\documentclass[preprint]{aastex}

\def\etal{{\it et al.}~}

\def\kms{\mbox{\,km\,s$^{-1}$}}

\begin{document}


\shortauthors{C\^ot\'e \etal }
\shorttitle{Star Formation in Centaurus A Group Dwarfs}


\title{Star Formation in Dwarf Galaxies of
the Nearby Centaurus A Group}

\author{St\'ephanie C\^ot\'e and Adam Draginda\altaffilmark{1}}
\affil{Canadian Gemini Office, HIA/NRC of Canada,
5071 West Saanich Rd., Victoria, B.C., Canada, V9E 2E7}
\email{Stephanie.Cote@nrc-cnrc.gc.ca}

\author{Evan D. Skillman}
\affil{Astronomy Department, University of Minnesota,
    Minneapolis, MN 55455} 
\email{skillman@astro.umn.edu}

\and 

\author{Bryan W. Miller}
\affil{AURA/Gemini Observatory, Casilla 603, La Serena, Chile;}
\email{bmiller@gemini.edu}

\altaffiltext{1}{Present address: Canada-France-Hawaii Telescope Corporation, 65-1238 Mamalahoa Highway, Kamuela, HI 96743; draginda@cfht.hawaii.edu}

\begin{abstract}
We present H$\alpha$ narrow-band imaging of 17 dwarf irregular
galaxies (dIs) in the nearby Centaurus~A Group. Although all large
galaxies of the group are or have been recently through a period of 
enhanced star formation, the dIs have normal star formation rates
and do not contain a larger fraction of dwarf starbursts than other
nearby groups such as the Sculptor Group or the Local Group.
Most of the galaxies in the group now have fairly accurately known distances,
which enables us to obtain relative distances between dIs and larger galaxies
of the group. We find that the dI star formation rates
do not depend on local environment, and in particular they do not
show any correlation with the distance of the dI to the nearest large galaxy
of the group.
There is a clear morphology-density relation in the Centaurus~A Group, 
similarly to the Sculptor Group and Local Group, in the sense that
dEs/dSphs tend to be at small distances from the more massive galaxies
of the group, while dIs are on average at larger distances. 
We find four transition dwarfs in the Group, dwarfs that show characteristics
of both dE/dSphs and dIs, and which contain cold gas but no current star formation. 
Interestingly the transition dwarfs have an average distance to the
more massive galaxies which is intermediate between those
of the dEs/dSphs and dIs, and which is quite large: 0.54$\pm$ 0.31 Mpc. 
This large distance poses some difficulty for the most popular scenarios 
proposed for transforming a dI
into a dE/dSph (ram-pressure with tidal stripping or galaxy
harassment). If the observed
transition dwarfs are indeed missing links between dIs and dE/dSphs,
their relative isolation makes it less likely to have been produced by 
these mechanisms. An inhomogeneous IGM containing higher density clumps 
would be able to ram-pressure stripped the dIs at larger distances
from the more massive galaxies of the group.
\end{abstract}


\section{Introduction}

The Centaurus~A Group of galaxies is one of the closest groups of
galaxies outside the Local Group (at a mean distance of about 3.8 Mpc).
Its composition is, however, much different from that of the Local Group, being
a heterogeneous assembly of early to late-type galaxies, and, in fact, it has
the largest dispersion of morphological types amongst all of the 55 nearest
groups reported by \citep{deV79}. Interestingly, it seems that all of its main
galaxy members have been through, or are experiencing, a period of enhanced 
star formation. The most prominent member of the group is Centaurus~A itself
(NGC~5128), a giant peculiar elliptical radio galaxy with powerful X-ray and
radio jets, which most probably suffered a recent major merger. Both M83 and
NGC~5253 are starburst galaxies \citep{cal99}; NGC~4945 has a Seyfert nucleus \citep{do96} 
and NGC~5102
is in a post-starburst phase \citep{da08}. One is then led to wonder if the group's dwarf
galaxy members also share this elevated level of activity. About 50 dwarfs
are now known in the group \citep{k07}, of which about two-thirds are gas-rich dwarfs
(dIs), and it would be interesting to know if these dIs also have an
elevated level of current star formation and a larger fraction of dwarf starbursts
than the dwarf galaxy population of less ``active'' groups, such as the nearby
Sculptor Group \citep{scm03a}. In other words, is the star formation activity
of the dwarfs influenced by their global group environment, or, does
it only depend on local conditions and internal processes?

A reasonable estimate of star formation rates (SFRs) in dwarf galaxies can be 
obtained by H$\alpha $ imaging, as it is known that in dwarf galaxies 
infrared and radio emission both can seriously underestimate the star formation rate
\citep[e.g.,][]{b03}. In the case of the Centaurus~A Group dwarfs, 
most of the dwarfs have relatively accurate distances \citep[e.g.,][with distances 
determined via the tip of the red giant branch method with errors typically of $\sim $0.5 Mpc]{k02}, 
and with optical diameters ranging from 1 to 10 arcmins 
they can have their full area imaged easily in one pointing.
An H$\alpha$ imaging survey can then produce much better star formation estimates,
especially compared to many galaxy studies at higher redshifts for which 
SFR estimates are derived from longslit data, for example \citep{bal97}, or even worse from
limited aperture fibers such as SDSS or 2dF surveys. As is obvious from
Figure 1 below, in some cases most of the star formation activity is found
in the outside regions, and a slit passing through the nucleus would severely
underestimate the true star formation rate. A detailed survey of HII regions
in each dwarf is also useful in order to find individual high surface brightness HII
regions required for follow-up spectroscopy for accurate abundance analysis 
\citep[e.g.,][]{scm03a, scm03b}. The vast majority of nearby dwarfs have only been
observed through longslits, not multi-object spectroscopy, so it is even more important
to know exactly how to position the slit to align them on the brightest HII regions,
from which one can reap the largest numbers of fainter emission lines.

Another useful outcome of a H$\alpha$ survey of the dIs members of a group is that 
one can identify so-called transition
dwarf galaxies, dwarfs that show characteristics in their morphology of both 
early-type dwarfs, dwarf ellipticals (dEs) or dwarf spheroidals (dSphs), and
late-type dwarfs (dIs). These transition dwarfs have cold gas but no (or extremely low) active
star formation so they will be mostly non-detections in the H$\alpha$ survey. These 
dwarfs have been proposed as a ``missing link'' between the two classes of dwarfs \citep[e.g.,][]{ggh03}.
The question as to whether they represent a real evolutionary link between 
dEs/dSphs and dIs is still controversial. In the Local Group there is a clear
morphology-density relation amongst the dwarfs, with dSphs being found predominantly
at short distances from large galaxies, while dIs are more spread out in the group.
Several scenarios have been proposed to efficiently transform a dI into a dSph
in the vicinity of larger galaxies. Did dSphs end up as they are
because of internal properties (from genetics), 
or are they dIs or pre-dIs that transformed into
dSphs simply because they happen to be at the wrong place at the wrong time? 
The study of transition dwarfs in groups might bring clues to their nature, 
and, indirectly, to the nature of dSphs too.

In \S 2 we will describe the H$\alpha$ observations, and \S 3 the SFRs 
for the Centaurus~A Group dwarfs will be derived. Section 4 will present the star
formation trends for dIs, its dependence on environment, and will discuss the
transition dwarfs in the Centaurus~A Group and other groups.

\section{Observations}

\subsection{Target Selection}

We observed all of the 19 dwarf irregular (dIs) galaxies members 
of the Centaurus~A Group listed in \citet{cfcq97}. 
Of these, 17 were detected,
and are listed in Table~\ref{tbl-1}. CEN5, which was also listed in \citet{cfcq97} as
a member dI,  was also observed but not detected,
and it has recently been confirmed to be instead a background galaxy 
\citep{bdj04}.
It is important for determining 
membership in nearby groups such as Centaurus~A to confirm the HI detections
of candidates dIs with an optical radial velocity, or obtain a rough distance
estimate through a color-magnitude-diagram (CMD) of the resolved stellar 
populations. This is necessary because of the possible confusion with 
Galactic High-Velocity-Clouds (HVCs) which can have 
HI velocities of several hundreds of km s$^{-1}$ (as was the case for CEN5).
Prior to our observations, all of the dwarfs listed in Table~\ref{tbl-1} 
had been confirmed 
to be bona fide members with optical spectroscopy, except  
UKS~1424-460 and UGCA~365 for which the memberships were confirmed
by tip of the red giant branch distances from HST by \citet{k02, k07}.

Since then, the total number of known dIs in Centaurus~A has grown
to 28 confirmed dIs and 4 more possible ones.
\citet{jfb98} found AM1318-444, not detected in HI but confirmed by optical
spectroscopy. \citet{b99}, as part of HIPASS, conducted a blind HI
survey of the group and proposed 10 new candidate dIs. 
The more recent HIDEEP survey \citep{m03} has revealed an additional possible dI member, 
J1337-33, and \citet{hke00} 
shows an HI detection for one more possible dI member, KK170. 
Although none 
of these last 12 HI detected dIs has been confirmed with optical spectroscopy,
ESO~321-G14, ESO~381-G18, IC~4247, AM1321-304 and HIPASS J1337-39, 
as well as HIPASS J1321-31, HIDEEP J1337-33 and ESO~269-G58 were all
followed-up with HST observations and all had their distances confirmed 
from their red giant branch tip \citep{k02, k07, g07}. 
Of these, only the first five appear to be
currently forming stars and to probably harbour HII regions, but unfortunately
were not observed in H$\alpha$ by us. Since then two of these, ESO321-G14 and 
AM1321-304, have been detected by Bouchard et al. (2009), and are included
in our Table 3 and subsequent Figures. 

Of the dIs that we observed (Table 1), a total of 14 now have measured 
distances based on the red giant branch tip \citep{k02, k07}.
For the three other dIs we assume a distance equivalent to the mean distance of
the Centaurus A Group of 3.8 Mpc, as determined by \citet{k07}.
In the nearby, southern Sculptor Group of galaxies,
it was found that galaxies seem to follow a distance-velocity relationship 
with very small scatter, defined with the measured distances to nine galaxies
\citep{jfb98},
of the form $v_{GSR}$(\kms ) = 119 $D$(Mpc) - 136. 
It was thus possible to assign a reasonable estimate of distance to a 
dwarf member based solely on its recessional velocity.
Unfortunately no such relationship seems to hold in
Centaurus~A. In fact, \citet{k07} have plotted over
50 galaxies in the region with both distance estimates (mostly from
red giant branch photometry) and recessional velocities, and it appears
that there are clearly two subgroups, with one subgroup of dwarfs
surrounding Centaurus A with a mean distance of 3.8 Mpc, and the other
around M83 at 4.8 Mpc, with no clean distance-velocity relationship.
While the Sculptor Group is a loosely bound group with no large disturbance from
the Hubble flow, the Centaurus~A Group on the other hand seems more evolved,
with more complex internal dynamics. 

The absolute magnitudes quoted in Table 1 are from the apparent magnitudes 
observed by \citet{c95}, corrected with updated Galactic extinction 
from \citet{sfd98} and converted to absolute magnitudes using the
adopted distances above.

\subsection{H$\alpha$ Imaging}

The H$\alpha$ images were obtained with the CTIO 0.9m telescope over 8 nights
in April 1999. The telescope was equipped with the Tek 2048 CCD, with a 
readout noise of 3.4 e$^-$ and gain of 1 e$^-$ ADU$^{-1}$,  
and a pixel scale of 0.396\arcsec pixel$^{-1}$. The observations usually consisted of 
$3 \times 1200$ sec through a narrow-band H$\alpha$ filter centred at 6559 \AA\ 
and with a FWHM of 64 \AA, followed by a $2 \times 1200$ sec for continuum 
off-band images through a filter centred at 6115 \AA\ and with a FWHM of 140 
\AA . R-band images were also acquired (typically 600 sec). The seeing 
varied between 1\arcsec\ and 1.4\arcsec ,
and the conditions were photometric for 6.5 nights out of 8.

\begin{figure}
\vbox to7.2in{\rule{0pt}{2.6in}}
\includegraphics{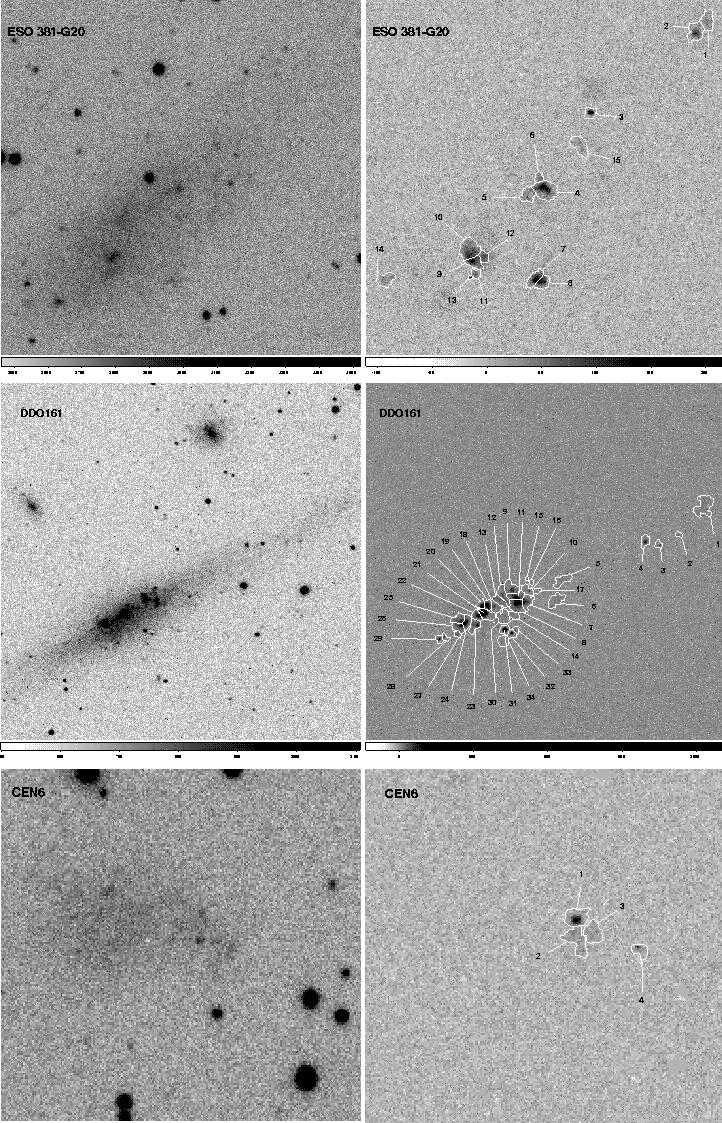} 

\figcaption{Images of 17 Centaurus A Group dwarf irregular galaxies.
The r-band images of the galaxies are shown in the left
panels and the continuum subtracted H$\alpha$ images are shown in
the right panels.  The HII regions are labeled and their fluxes
are listed in Table 2.
The field of view is 150\arcsec\ $\times$ 150\arcsec\ , except for
Cen6, NGC~5237, IC~4316, ESO~222-G10, ESO~272-G25 (75\arcsec\ $\times$ 75\arcsec\ ),
ESO~325-G11, ESO~383-G87, ESO~223-G09 (300\arcsec\ $\times$ 300\arcsec\ ), DDO~161
(330\arcsec\ $\times$ 330\arcsec\ ), and ESO~274-G01 (540\arcsec\ $\times$ 540\arcsec\ ).
SEE BETTER FIGS IN AJ PAPER\label{fig1}}
\end{figure}

The data reductions were performed mostly with IRAF following the usual 
procedures. The different images in each filter were registered and then 
co-added. The co-added off-band image was smoothed with a Gaussian so that 
in the final image the point-spread functions matched as closely as possible
those in the H$\alpha$ image. This continuum image was then scaled 
appropriately using half a dozen isolated bright stars and then 
subtracted from the H$\alpha$ image. Figure~\ref{fig1} shows the R images
and the final continuum-subtracted H$\alpha$ images. Table 2 lists
the positions and fluxes of all the HII regions found in each dwarf galaxy.
HST Guide star Reference Frame scans from the STScI Digitized Sky Survey
were used to derive accurate positions for the HII regions.

The HII regions are rarely isolated, but are most often located in large complexes
and occasionally exhibit complicated morphologies with loops or 
filaments, which would make it difficult for an automatic procedure to 
delineate them. Therefore it was necessary to determine the boundaries of
each HII region by eye, and, using POLYPHOT, the fluxes were obtained by 
integrating all the emission within the regions. 
Each distinct emission peak was defined to be a separate HII region,
ensuring that each of them correspond to a separate excitation source.
It may be that some of these peaks are just density peaks in the excited gas
rather than genuinely separately excited HII regions. In addition, 
there might be some degree of randomness
in the way adjacent HII regions are separated, but these
difficulties would also affect regions found by automatic software methods.
However, it should be kept in mind that these difficulties will have an 
influence on the faint-end of the luminosity function of HII regions.
When filaments or extended diffuse emission were present, they were 
included in the most nearby HII regions, since most probably they are ionised
by escaping photons from these regions and do not have their own ionising 
stars. In some cases, some diffuse emission regions were found fairly 
isolated in the galaxies and, hence, were counted on their own.
Most of the sample dwarfs did not have significant diffuse emission,
just as the faint dwarfs (M$_B >$-15.9) of \citet{shk91}.
The borders of the HII regions
were set to a constant H$\alpha$ surface brightness level for each galaxy,
at our detection limit estimated at about $\sim 3.0 \times 10^{-17}$ 
ergs cm$^{-2}$ s$^{-1}$ arcsec$^{-2}$. Errors in fluxes are typically around
10\%, mostly due to absolute flux calibration uncertainties. No correction 
for [NII] contamination has been applied to these
fluxes; dwarf galaxies are known to have very low nitrogen abundances
\citep[see, e.g.,][]{scm03b} so this introduces an additional $\sim $
6\% flux uncertainty at most.
H$\alpha$ luminosities were calculated from the H$\alpha$ fluxes using
the distances of Table 1 and assuming a Galactic extinction correction of
the form $A(H\alpha ) = 2.32 E(B-V)$ \citep{mh94}, with reddening
values from \citet{sfd98}. 
Almost all of our galaxies (except Cen6, UGCA365 and NGC5237) have previous 
H$\alpha$ measurements in 
the literature, mostly from the recent survey of \citet{kl08}. 
Our fluxes agree within 50\% overall with those of this survey. 
They agree better with those of other authors:
IC~4316 and NGC~5408 were imaged in H$\alpha$ by \citet{gh87},
who found total fluxes of 1.3 and 34 $\times 10^{-13}$ erg 
s$^{-1}$ cm$^{-2}$, agreeing well with our total fluxes of 1.2 and 39.2 
$\times 10^{-13}$ erg s$^{-1}$ cm$^{-2}$. DDO161 was observed by \citet{meu06}
who gives it a SFR of 9.77$\times 10^{-3} M_{\odot} yr^{-1}$, very close to our 
1$\times 10^{-2}$. Finally \citet{kai07} obtained for UKS1424-460 a SFR
of 1.3$\times 10^{-4} M_{\odot} yr^{-1}$, again very close to our 
1.4$\times 10^{-4}$.

\section {The HII Regions and the SFRs of the Centaurus~A Group dI Galaxies}

\subsection{The HII Region Distributions and Luminosities}

The HII regions in the Centaurus~A Group dwarfs are mostly distributed asymmetrically
throughout the galaxies, as is seen in the Sculptor Group dwarfs \citep{scm03a} 
and other nearby dIs \citep{bha98, vz00}.
This is the case both for the dIs with very little star formation
(e.g., ESO~324-G24), as well as the dIs with large numbers of HII regions 
(e.g., DDO~161). In brighter spiral galaxies, HII regions will
most likely be concentrated in the inner parts, and they typically have H$\alpha$ 
surface brightness profiles similar to broadband V or R surface brightness profiles,
peaking in the center \citet{kk06}. 
Our Centaurus~A Group dIs have  
their brightest HII regions most often in the outer parts of
the galaxy, sometimes out to the very edge of the optical disk, which is often
the case for dIs \citet{bha98}. Note that this is true even when accounting for the fact
that there is more area at large radii. Only two
of the dwarfs have their dominant brightest HII region centred in the
nucleus: DDO~161, which is one of the largest dwarfs in the sample, and
UKS~1424-460, which has only one HII region. One can see easily from
Figure~\ref{fig1} why longslit spectroscopy of dwarfs to estimate their SFRs
is rather inefficient, like what is routinely done in surveys at higher redshift.
A slit positioned along the major axis of the galaxy would miss the
brightest HII regions in many cases, and seriously underestimate the 
total H$\alpha$ flux for the majority of dwarfs. In some cases (e.g., ESO~324-G24)
a slit with a typical slitwidth of 1 arcsec would  miss all 
the HII regions, even if the dwarf was at at a distance up to $\sim $ 250 Mpc 
(where 1 \arcsec $\sim$ 1 kpc). In fact \citet{pg03} have compared H$\alpha$ fluxes 
determined by spectroscopy with those measured on H$\alpha$ images for a sample
of emission-line objects at a mean redshift z=0.026 and found that the spectroscopy
yielded fluxes only 1/3 of the total emission-line flux.

Figure~\ref{fig2} is a histogram of the HII region luminosities for all
the Centaurus~A Group dwarfs. The peak is at roughly 10$^{37.4}$ erg s$^{-1}$, 
most of the HII regions being rather low luminosity, which is similar to what
is seen for the Sculptor Group dwarfs as well as other nearby dwarfs.
The fact that the Centaurus A Group contains on average a significantly larger number of brighter dwarfs
than the Sculptor Group (Figure~\ref{fig2}) does not translate in a significant
differences in the overall HII regions luminosities distribution.
Figure~\ref{fig3} shows the fraction of the total H$\alpha$ luminosity in 
each galaxy that is contributed by HII regions of a particular luminosity range,
ranging from regions with luminosities like that of the Orion nebula
$\sim 10^{37}$ erg s$^{-1}$ to supergiant HII regions with $L\geq 10^{39}$
erg s$^{-1}$. Despite the fact that the Centaurus~A Group dwarfs contain many
more HII regions than those in the Sculptor Group, it appears that the luminosity 
distributions of these HII regions are very similar for the two groups.
The bulk of the HII luminosity in these dwarfs comes from regions with
modest luminosities $10^{37}-10^{38}$ erg s$^{-1}$. The only Centaurus~A Group
dwarf
which contains supergiant HII regions is NGC~5408, classified as a starburst
by \citet{khdl01}, in which Wolf-Rayet features were also detected
\citep{scp99}. The percent contribution from the supergiant regions
to its total HII luminosity is just above 70\%, which is typical for
Blue Compact Dwarfs and starbursts \citep{yh99}.
This dearth of supergiant regions in the dwarfs is common, for example in their
sample of 29 nearby normal dIs Youngblood \& Hunter (1999) found only 2 with supergiant regions.
For this reason the fraction of
H$\alpha$ emission contributed by all of the dwarfs in the Centaurus~A Group
compared to that from the main galaxies is negligible: their total
contribution comes to less than $1\times 10^{-11}$ ergs cm$^{-2}$ s$^{-1}$,
which is less than the H$\alpha$ emission from the three largest supergiant
regions alone (amongst many) of M83 \citep{rk83}.  

Only two dwarfs out of 19 listed as 
Centaurus~A Group members in \citet{cfcq97} were not detected here 
in H$\alpha$: UGCA~319 (also known as SGC1259.6-161), and NGC~5206. 
In the case of the latter this is not surprising, because it is a dE type,
classified as T=$-$3 in the RC3 and T=$-$2 in \citet{lv89},
with a well-defined $r^{1/4}$ profile \citep{pbka93}.
It was not detected in HI (down to 7.8 $\times$ 10$^6$ M$_{\odot}$) 
and its velocity comes from optical absorption lines
measurements \citep{cfcq97}. UGCA~319 on the other hand has 
now been detected at the Mount Stromlo and Siding Spring 2.3m
by \citet{bou09}, with a flux of 1.6$\times 10^{37}$ erg\  s$^{-1}$.
 
\subsection{Global Star Formation Rates and Timescales}

H$\alpha$ luminosities were converted to current SFRs as in
\citet{ktc94},  with:
\begin{equation}
SFR(total) = {{L(H\alpha )}\over{1.26\times 10^{41} erg\  s^{-1}}}\
 M_\odot\  yr^{-1}
\end{equation}
which has been derived for normal spiral galaxies with a Salpeter IMF, 
and is the conversion adopted by \citet{k98} in his grand synthesis of
global SFRs in galaxies.  
No corrections were made 
for internal extinction \citep[as had been done for spiral galaxies in][]{k98}
since extinction is normally quite small in these low metallicity systems.
A possible exception would be dwarf starbursts, but in their case it is 
difficult to quantify \citep[see, e.g.,][]{cdsbcm03}. 
Adopting a single conversion factor from H$\alpha$ luminosity to SFR 
for dwarf galaxies which has been derived for normal spiral galaxies certainly
carries some uncertainty.
One can think of several possible biases, primarily
how the IMF might not be universal and/or the production of ionising photons 
by the stars 
might be metallicity dependent. It seems indeed that the IMF in low-luminosity
galaxies has fewer massive stars (from a large SDSS study from Hoversten \& Glazebrook 2008),
either by a steeper slope or lower upper mass cutoff. We nevertheless apply 
this conversion
factor to our dIs for consistency so that our sample can be compared 
directly to other previous studies which have all used this same factor.
One should keep in mind though that SFR levels below about 
$10^{-3}$ $M_{\odot}$ yr$^{-1}$ might be inaccurate, since it seems from simulations
that it is at 
this level that stochastic effects come into play and a depleted upper mass end
of the IMF result in a lack of massive stars responsible for H$\alpha$ \citep{tre07}.

The SFRs listed in Table~\ref{tbl-3} were calculated using equation (1) 
with H$\alpha$ luminosities obtained by summing 
all the HII region luminosities listed in Table~\ref{tbl-2}.
Most of these SFRs are very low relative to ``normal spiral galaxies'', with values
ranging from $3.1\times 10^{-5}$ up to $3.4\times 10^{-2}$ $M_{\odot}$ yr$^{-1}$, 
with the 
possible exception of NGC~5408 at $8.8\times 10^{-2}$. From the compilation of
150 galaxies of the Local Volume by Karachentsev \& Kaisin (2007), spirals have SFRs ranging
from $1\times 10^{-1}$ up to $5.5$ $M_{\odot}$ yr$^{-1}$.
 However, these dwarfs SFRs 
are comparable
to the low levels of star formation typical of dIs (typically $1\times 10^{-3}$ 
$M_{\odot}$ yr$^{-1}$ for a $M_B=-14$ dI in Karachentsev \& Kaisin 2007). They are slightly
larger on average than those obtained for the Sculptor Group dIs, but
this is normal since the SFR is known to be a function of absolute magnitude,
and the Centaurus~A Group dIs sample includes more brighter dwarfs 
than the Sculptor Group. 
In Figure~\ref{fig4} we have plotted the SFRs normalised to L$_B$ versus
versus M$_{HI}/L_B$, and it can be seen that 
the Centaurus~A Group dIs cover a range comparable to that of the 
Sculptor Group dIs or the Local Group dIs. 
What is noticeable in this Figure 
is that there seems to be a grouping of Centaurus~A Group dIs at low
M$_{HI}/L_B$, compared to other dIs. This is probably
due to the fact that M$_{HI}/L_B$ scales with absolute magnitude
(fainter dwarfs being more HI rich proportionally), and so the brighter 
Centaurus~A Group dIs
would have lower M$_{HI}/L_B$ than the average dIs. Indeed, in this group
four of the five dIs are among the very brightest dIs of the Centaurus~A Group.
What also stands out in Figure~\ref{fig4} are
the Local Group objects at high SFR$/L_B$ 
(like NGC~6822 and IC~10). Apparently it is the Local Group dIs which have
unusual SFR properties compared with other groups' dIs. The only
dI of the Centaurus~A Group that qualifies as a starburst, NGC~5408, is indeed the one
with the highest SFR$/L_B$, at the top end of the distribution. There are no strict definition of starburst on which all astronomers agree, although commonly used denitions are that:
a) continued star formation with the current SFR would exhaust the available gas reservoir
in much less than a Hubble time; and b) the current star formation normalised by the
past averaged SFR is much greater than unity \citep{ga05}. A definition recently
suggested specifically for dwarf starbursts is that their integrated H$\alpha$ equivalent width
should be above 100 \AA \citep{lee09}. NGC~5408 satisfies all 
of these criteria, and is the only dwarf of the group to do so.  It thus seems
that the Centaurus~A Group dwarf population does not follow the enhanced 
star forming activity experienced by its larger galaxies. The average SFRs
of the dwarfs, as well as the number of dwarf starbursts,
are all consistent with what is seen for other nearby dwarfs. Whatever
has triggered the star forming episodes of the larger members of the
group does not seem to have had any obvious effect on the nearby dIs. 
With only one dwarf starburst amongst the $\sim $ 30 dIs known in the 
Centaurus~A Group, these numbers are consistent, but on the low side, of what is found
overall for nearby dwarfs: Lee (2006) in an 11 Mpc H$\alpha$ UV survey 
finds that 6$\pm$ 3 \% of low-mass galaxies are currently experiencing
a starburst. This is despite the fact that the Centaurus~A Group
has been carefully combed in HI so that compact HII galaxies would not have been
missed, contrary to the field in the 11 Mpc volume. 

Table~\ref{tbl-3} provides two interesting timescales for our dIs 
to estimate the significance of their SFRs. The first one is $\tau _{gas}$,
the gas depletion timescale (= total gas mass/SFR), which is an estimate
of the number of years a galaxy may continue to form stars at the current
rate until gas depletion. The other one is $\tau _{form}$, the star formation 
timescale, which is the ratio of the mass of stars present to the current rate
of star formation. This provides a rough indication if the galaxy is 
currently forming stars at a lower or higher rate than it has in the past. 
Note that $\tau _{gas}$ is a lower limit since it does not account for
all the material that will be recycled over the course 
of normal stellar evolution. Despite this, the values of 
$\tau _{gas}$ for the Centaurus~A Group dIs are very large, several times the present
age of the universe, except for a few particular exceptions. Obviously
galaxies which are starbursting will exhibit much shorter gas consumption
timescales, and this is the case for NGC~5408. The only other two objects with
very short gas depletion times are the two oddest Centaurus~A Group's galaxies:
NGC~5237 and ESO~272-G25.
NGC~5237, having the appearance of a starburst galaxy but with a nicely elliptical shape,
has been morphologically typed from T=$-$5 \citep{vvi74}
to T=$+$5 \citep{ph98}. \citet{t92} has suggested
that it is the remnant of the spiral which collided with Centaurus~A (NGC~5128)
and consequently lost half its disk material (creating the ring of gas and 
dust prevalent in Centaurus~A) and that the rest was ejected to form this 
object out of the dusty, gas-rich remaining disc material. It certainly has
many features of early-type dwarfs: in addition to its morphology, its
M$_{HI}/L_B$=0.17 is typical of dEs and its color B-I=1.6 \citep{c95} 
is the reddest of all the Centaurus~A Group dI candidates. On the other hand, it
seems to be rotating too fast for a dE and more in line with a dI of this
magnitude. No rotation curve is available but the Parkes global HI profile
has a width at the 20\% level of $\Delta V_{20\%}$=89 \kms \citep{cfcq97}.
It might be indeed that NGC~5237 is undergoing a transformation from
stripped late-type spiral to a dE.  
The other unusual object is ESO~272-G25, classified as 'peculiar' by 
\citet{l84}.
The odd thing about ESO~272-G25 is that it was not detected in HI,
down to 7.8$\times 10^6$ M$_\odot $. This is a factor of $\sim$ 5 times lower
than what would be expected for a dI of this luminosity. It seems to contain
some bright emission knots surrounded by a low surface brightness envelope,
reminiscent of some Blue Compact Dwarfs (BCDs). However, BCDs have normal 
M$_{HI}/L_B$ \citet{hkp05}. It has several chains of HII regions,
and the morphology of the underlying envelope is not dE-type. One might
argue that it is a recently stripped dI, but ESO~272-G25 is situated in
the periphery of the Centaurus~A Group, far from any of the massive members. 
Another way of making the HI disappear is if the galaxy recently underwent 
a large burst or several bursts of star formation which would have consumed 
most the gas available. Normally one would not expect the efficiency to be so high,
and besides its SFR/$L_B$ is rather ordinary. ESO~272-G25 is reminiscent of
POX~186, another BCD with no HI detected \citep{bc05}.  In this case, it
has been hypothesised that the present burst of star formation has ionised
most or all of the cold gas.  Both NGC5237 and ESO272-G25 would need
some dedicated follow-up observations to understand their true nature.

Figure~\ref{fig5} shows histograms of $\tau _{gas}$ for the Centaurus~A Group
dIs compared to the Sculptor Group dIs \citep{scm03a}, the Local 
Group dIs \citep{m98}, and the large sample of isolated dIs from \citet{vz01}.
It is the norm for dIs to have large values of $\tau _{gas}$. In comparison 
the large sample of spirals of \citet{ktc94} have a mean $\tau _{gas}$ of
3.6 Gyr (log($\tau _{gas}$)=9.6). It thus appears that dIs can continue
to form stars with their current level of low SFRs in a
continuous manner for several Hubble times without running out of fuel.
\citet{ktc94} pointed out that in calculating $\tau _{gas}$ the total gas mass detected
in the galaxy is used when in fact one might want to consider only
the gas within the optical radius of the galaxy, which would be effectively
available for star formation. Only a handful of dwarfs here have HI aperture synthesis 
data such that this could be evaluated, but it turns out that from the HI  
radial surface density profiles of C\^ot\'e et al. (2000) the extended gas represent 
only between 5 to 10\% of the totalHI mass, because the gas profiles drop off sharply
beyond the optical edge. 
All of the dIs in Figure~\ref{fig5} had 
values of $\tau _{gas}$ calculated in the same way, using all the gas, for comparison 
purposes. It appears that the Centaurus~A Group dIs and the Sculptor Group dIs have a
similar distribution of $\tau _{gas}$, and that it is the Local Group dIs
that have a different distribution, with fewer galaxies with long $\tau _{gas}$
which could continue to form stars at the present rate for a long time, 
and more galaxies at small $\tau _{gas}$ which will be depleted soon.

The same thing is seen in the histograms of the star formation timescales
$\tau _{form}$ for these same groups of dIs (Figure~\ref{fig6}). 
The values of $\tau _{form}$ calculated
here are just rough estimates; what one really would want to evaluate is
the ratio of the past average SFR to the current SFR. However, to calculate
the past average SFR one would need to know the mass of stars formed over
the lifetime of the galaxy. This means some assumptions need to be made
about the mass-to-light ratio of the stellar material, and also about
the age of the galaxy. We have derived our estimates following the
simplification of \citet{h93}, who simply adopts a mass-to-light ratio 
of one (which makes sense for typical colours of dIs). With M/L = 1,  
$\tau _{form}$ becomes simply equal to L$_B$/SFR. In Figure~\ref{fig6},
similarly to Figure~\ref{fig5}, the Centaurus~A Group dIs and Sculptor Group
dIs follow the same distribution while the Local Group dIs distribution
stands out, with more dIs with short $\tau _{form}$, again because of
the higher number of dIs starbursting or forming stars at a much higher
rate than they have in the past. Using the \citet{k83} sample of galaxies,
\citet{h93} obtained average $\tau _{form}$ values of 60 Gyr for early-type
spirals, 15 Gyr for late-type spirals, and 8 Gyr for irregulars. Here
the dIs of nearby groups have slightly longer $\tau _{form}$, which means
that they have lower current SFRs, about sufficient to 
build the current stellar population over a Hubble time.  

A more direct way of evaluating the star formation history of a galaxy
is through color-magnitude diagrams (CMD) of their resolved stellar populations.
The numbers of stars in different regions of the CMD (main sequence, 
red giant branch, etc.) can be used to retrace the star formation history.
CMDs of many of the Centaurus~A Group dIs were built from HST/WFPC2 data
by \citet{k02}.  Unfortunately the data are mostly too shallow
for in-depth star formation history reconstruction (their goal was to 
determine distances using
the tip of the red giant branch). One can distinguish in most
of the dIs a main sequence indicating recent star formation, as well
as an important intermediate age component, which is at least compatible
with deep CMDs obtained for Local Group dIs \citep[e.g.,][]{dp98},
showing that, except for a few starbursty dwarfs,  
star formation has occurred not in (a few) big bursts but rather
in a continuous stochastic manner, with slow and steady star formation
through time with some periods with a slightly more elevated rate (by a 
factor of only a few).

\section{Star Formation Trends in dIs}

The Centaurus~A Group dwarfs show a very wide range of SFRs,
as seen from Table~\ref{tbl-3}, even when scaling by the luminosity,
as in Figure~\ref{fig4}. The SFR/$L_B$ in Centaurus~A Group dIs and other
local dIs range roughly from $\sim 10^{-12}$ to $10^{-8} 
M_{\odot}$ yr$^{-1} L_{\odot}^{-1}$, overlapping with that of normal spirals
which have typically SFR/$L_B$ $\sim 10^{-9} M_{\odot}$ yr$^{-1} L_{\odot}^{-1}$.
One then must wonder if the star formation activity of a particular galaxy
really depends in any way on some of its global properties, such as
gas-richness, color etc.\ (\S 4.1). In \S 4.2 we will look into
the local conditions that seem necessary for star formation to take place,
and then in \S 4.3 we will inspect
if the star formation activity is perhaps more dependent on the
environment in which the dwarf galaxies are located, both locally and 
globally.\S 4.4 will discuss the morphology-density relation in the group,
and \S 4.5 will offer a global comparison between the Centaurus~A Group
and the Local and Sculptor Groups.

\subsection{SFR Dependence on Global Properties}

We will investigate here if the SFRs correlate in any way with some
global parameter of the galaxies, such as magnitude, central surface brightness,
color or HI content, as was done in Hunter et al. (1982), van Zee (2001) and 
Hunter \& Elmegreen (2004). 
To look at how SFRs might depend on these global
parameters, one must first somehow normalise these rates to some
measure of the size of the galaxy -since obviously larger galaxies
have a larger volume to harbour HII regions. One may think then that
the best normalisation should be achieved by looking at SFR per unit
mass of the galaxy. As a measure of the mass, one can use a) the
total baryonic mass, e.g., the mass in stars and gas, or b) the total
dynamical mass. The difficulty with the first option is that, although
all the dIs in the sample have good HI mass estimates, the mass of 
their stellar disk can only be estimated approximately because of
the unknown mass-to-light ratio for the stellar material \citep[at best, some
estimates could be obtained using the color indexes of the galaxies, e.g.,]
[]{lea06}.
The second option is not much better, since to calculate the dynamical
mass $M_{dyn} = R V_{rot}^2/G$, one needs the rotation velocity of the
dI. Some of the dIs in our sample have been mapped in HI with aperture 
synthesis and have reliable rotation curves so these $V_{rot}$ are known
with some accuracy. But for the majority of the sample only estimates
of $V_{rot}$ are available based on single-dish HI width measurements 
and rough inclination angles (and $M_{dyn}$ is inversely proportional to
(sin$i$)$^2$). For these reasons it is safer to normalise the SFRs
to some sort of measure of the area of the galaxy.
\citet{hg86} chose to use the Holmberg radius 
(the radius at which $\mu _B$=26.6 mag arcsec$^{-2}$) 
for their sample of Irregular galaxies. However, our sample is 
composed entirely of dIs that are typically of lower surface-brightness 
(in fact, on average, surface brightness correlates with size).
The choice of the Holmberg radius
is therefore not appropriate, as the lowest surface brightness dwarfs
profiles will reach the Holmberg radius 
much sooner. Consequently a unit in terms of the
luminosity exponential profile scalelength is chosen, and the final 
normalisation adopted is an area calculated as $\pi (1.5 \alpha)^2$
(the Holmberg radius corresponding roughly to 1.5$\alpha$ 
for our range of dwarfs). 

In Figure~\ref{fig7}, SFR/area is plotted 
against several global properties: the absolute Blue Magnitude,
the B central surface brightness, the HI mass to luminosity ratio $M_{HI}/L_B$,
and the color $B-R$. No clear correlations are present for any of
these parameters, confirming the results of previous studies, e.g. van Zee (2001)
and Hunter \& Elmegreen (2004). 
The normalised
SFR versus the surface brightness shows at best a hint of a correlation,
but this is not surprising, as an enhanced SFR would bring as a normal consequence
the brightening of the surface brightness. There is also a hint of a trend
for lower luminosity galaxies to have lower SFR, but the scatter is very large
in SFR. There is no trend between SFR and colours (B-R), even though bluer
colours are a sign of a stellar population with a younger mean age.
For our set of dwarfs it is therefore not a global galaxian parameter 
that clearly influences the strength of the
star formation in a given object. It is perhaps more likely that 
it is the local conditions that are determining the level of star formation 
activity in the galaxy.

\subsection{SFR Dependence on Local Conditions}

It has been observed in numerous dIs that the majority of their HII regions,
especially the brightest ones, seem to be associated with local peaks in the
HI distribution, most often being located along ridges of regions of higher
HI column densities. In spiral galaxies star formation seems largely regulated
by the propagating spiral density waves, but dIs, except for the largest ones,
do not have the required gravitational potentials to support these. Successful star  
formation `laws' have been devised for spirals, relating locally the star 
formation rate to a power of the gas density, such as in the `Schmidt law',
or the now more popular \citet{k98} version. Moreover there is good evidence
that there exists some threshold below which the star formation plummets abruptly. 
The \citet{t64} dynamical threshold \citep[see also][]{k89} is based on the idea
that there should be a critical surface density above which a self-gravitating
infinitely thin rotating gas disk is locally unstable to axisymmetric perturbations; 
this critical surface density being proportional to the gas velocity dispersion 
and the epicyclic frequency. However, it has often been observed in dIs that
their gas densities are well below their threshold for star formation across
the entire galaxy, and yet they contain many star forming regions.

It has been suggested that observations could be explained instead by a star 
formation rate dependence on gas volume density (rather than surface density), 
which diminishes drastically in the outer disk
due to the vertical flaring of the gas layer, and this would remove the need
for a large-scale gravitational threshold \citep{f02}. DIs' star-forming properties
might be difficult to explain with this scenario though, as they are more puffy
than larger spirals, and have proportionally more flaring, so one would expect
lower SFRs throughout the dwarf compared to a spiral judging by
the gas surface density, which is contrary to the observations that dwarfs observed
SFRs are much higher than predicted. In dIs, regions of high SFRs 
are popping everywhere in the galaxy, very often at larger radii while
  none are observed in the central regions. Therefore the argument
  of gas 'dilution' as the gas enveloppe flares at larger radii and therefore
  the gas volume density lowers does not agree well with the observed SFRs.
The newly proposed model of thermo-gravitational instability of \citet{s04} is 
a promising alternative. According to \citet{s04} the UV background radiation 
implies a surface density threshold for the formation of a cold gas phase
(and this threshold agrees with the observed one). The transition to a cold phase,
associated with a drop in the pressure, triggers gravitational instabilities and
hence star formation. For gas densities below the threshold, 
self-gravitating gas clouds
are kept warm and stable by the UV background radiation.
However, it seems that somehow there must be a feedback between the energy 
injected in the
ISM (through SN, stellar winds etc) and the ISM itself to regulate the star
formation in dIs. \citet{hi00} argues that the heating (from stellar feedback)
must be very efficient in dIs because of their small size, and on the other hand
the cooling does not become effective because of their low metallicity abundances.  
With the balance of these two processes the intermittent 
star formation activity of small-size dIs can then be reproduced.

Looking at the sample of Centaurus~A Group dwarfs addressed here, only five
dwarfs have been properly mapped in HI by aperture synthesis:  
ESO~381-G020, DDO~161, ESO~444-G084 and ESO~325-G011 
from \citet{ccf00}; and IC~4316 from \citet{dzdbf02}, 
all done with the ATCA. Those first four dwarfs 
all have gas surface densities lying well below the critical density
for star formation \citep[calculated following][]{k89} as shown in 
\citet{ccf00}. Despite this they all harbour numerous star
formation regions as we have seen above. In each case the HII regions
are found to be contiguous with the HI density peak or circling very closely
the region of the HI peak, as is often observed in dwarf galaxies
\citep[e.g.,][]{s88, t94}. 
When there are two distinct separate HI peaks,
the bright HII regions are even seen in two clusters, surrounding each of the 
peaks. This is similar to what was seen in the dwarf galaxies of van Zee et al. (1997a),  
see in particular UGC5764 in their Figure 11. In each of these four dwarfs
it is found that the sites where current star formation activity is taking
place are always in regions where the HI gas column density is at least 
7.3 $\times 10^{20}$ atoms cm$^{-2}$ (including all outlying HII regions,
down to our detection limit). And the brightest HII regions in each dwarf 
are limited to
regions of over 9.8 $\times 10^{20}$ atoms cm$^{-2}$. For IC~4316 it is
found that the HII regions lie within the contour of the 2 $\times 10^{20}$ 
atoms cm$^{-2}$ level, however the spatial resolution for the IC~4316 data 
is twice that of the \citet{ccf00} data, with a synthesised beam
corresponding to about 1 kpc at the distance of the dwarfs. Conversely
regions with HI gas column density of at least 7.3 $\times 10^{20}$ 
atoms cm$^{-2}$ do not necessarily exhibit fresh HII regions, so this level
of gas surface density seems like a necessary condition for active star
formation in dwarfs but is not the sole one. Since these regions are very
much near the central parts of the galaxies it is unlikely that there are
large volume density variations due to e.g., flaring in these parts of
the galaxies. It thus seems that a simple gas volume density limit is not
sufficient to explain star formation activity. However in the inner parts
of the galaxies one does not know the detailed heating processes and the
exact locations where the stellar events have occurred which will have 
influenced the temperature and density of the gas. On the other hand the 
cooling rate, dependent on the metallicity, should be fairly uniform
across the dwarf, since the metallicity does not show strong gradients
in cases where several HII regions were measured in the same object. It is
thus very possible that the balance of the two, heating and cooling, indeed
produces the star formation activity observed. Note that in dwarf galaxies
it is very probable that there is still, as in larger galaxies, some low
level of star formation beyond these thesholds, as can be traced now in
the UV (\citep {boi07}). These thresholds seem to be where H$_\alpha$ 
is no longer detected because the number of massive stars ($M > 10 M_\odot$)
with large inonizing fluxes responsible for it are rapidly dwindling, while the
UV continuum emitted by slightly smaller, longer lived stars can still be
observed.

\subsection{SFR Dependence on Environment}

There are many reasons to suspect that a dwarf galaxy's local 
environment might play a role in increasing/decreasing its 
star formation. Depending on the local galaxy density, e.g., if the
dwarf is rather isolated or in a group or in the middle of a cluster,
various processes are expected to affect its star formation. 
Star formation might first be induced as a dwarf starts approaching
the cluster centre, due to the pressure in the intracluster medium 
leading to compression of gas clouds, or cloud-cloud collisions
\citep{e97}. Then ram-pressure stripping might remove its
reservoir of gas, therefore halting its star formation activity.
Strangulation, where hot gas is depleted from the dwarf's halo after
it enters a hot medium (which means no more hot gas can cool and
eventually form new stars) is another possibility \citep{ltc80}.
It has been argued though that these processes are unlikely
to be very effective in a group environment: a) the pressure force
for the stripping depends on the square of velocity dispersion of the group
and these are too low in groups ($\sigma < 400$ km s$^{-1}$) to produce a 
significant effect; b) strangulation is a slow process, which takes $\sim$ 1
Gyr in clusters, and moreover it requires an intracluster medium 
\citep{c06}. Galaxy-galaxy interactions, on the other hand, are
frequent in groups, and this is a process that should induce star 
formation in the participating galaxies, should they be mergers
or just low velocity interactions.

Hunter \& Elmegreen (2004) investigated how the SFRs of galaxies depend on 
their proximity to other galaxies, and found no correlation 
in their sample of 94 nearby galaxies.
However the distances used for the galaxies were for the great majority only 
estimated from Hubble's law. 
Even in superclusters, like Virgo and Coma, \citet{g98, g02}
found that the star formation rate for late-type galaxies are the same 
as in the field. They found decreasing SFRs with
decreasing distance from the cluster centre only for bright galaxies,
but not for the dwarf galaxies. In compact groups too \citet{iv99} 
found the same median SFRs in the middle of the
groups compared to the field for their sample of disk galaxies.

In Figure~\ref{fig8} we explore how the SFRs of our dIs vary depending
on the dwarf's distance to the nearest spiral galaxy.
Only the Centaurus~A Group dIs  
for which there is an accurate distance measurement were used, 
and ESO~223-G09 was excluded too because at a distance of 6.4 Mpc
it has to be considered a background object. 
The plot show the normalised SFR/area versus the distance of
the dwarf to the nearest large galaxy of the group. 
As expected, there is no correlation, either for the Centaurus~A Group dIs, 
or for the combined sample with the Sculptor Group dIs and the Local Group dIs. 
Similarly, if one looks
at the SFR/area versus the distance to the group centre there is no
apparent trend. It thus appears at first sight that the star formation activity of a dI in 
nearby groups does not depend on its immediate local environment.
However the existence of a morphology-density relation in both the Local Group
and the Sculptor Group (\citep{scm03a}) points to some underlying environmental
effects at play, and below we take a look at the situation in 
the Centaurus~A Group.

\subsection{The Morphology-Density Relation for the Centaurus~A Group Dwarf Galaxies}

Cluster galaxies have long been known to follow a morphology-density
relation, where early-type galaxies (including early-type dwarf galaxies)
are predominantly found in the higher density regions of the cluster whereas
the late-type and irregular galaxies are in the periphery in lower
density regions. This morphology-density relation is also found to work
in the Local Group, in which faint dSphs and dEs
are found predominantly in the vicinity of the Milky Way and M31 while
dIs are widely spread in the group \citep{vdb94a}.
This is also true in the nearby Sculptor Group, where the 
majority of the dwarfs now have good distance estimates and therefore a 3D
picture of the group is possible, and there too one finds the early-type
dwarfs at a mean distance of only 0.22 $\pm$ 0.21 Mpc from the nearest spiral  
galaxy while dIs are at a mean distance of 0.95 $\pm$ 0.61 Mpc \citep{scm03a}. 
Interestingly, the so-called `transition' dwarf galaxies,
objects that show characteristics of both dEs and dIs, are found
at intermediate distances,
with a mean distance of 0.50 $\pm$ 0.34 Mpc to the nearest
(large) spiral galaxy \citep{scm03a}. Naturally, one is then led to think that there
are probably some environmental effects at play which are able to drive
a normal dI into a transformation
into a dE/dSph, as the dI falls into
the group potential. Possible such effects include ram-pressure stripping
\citep{gg72}, galaxy harassment, which transforms a small
disk galaxy into a left-over dE or dSph \citep{mkldo96}, or tidal
stirring, where repeated tidal shocks partially strip the halo and disk
of a dI and reshape it into a dE/dSph \citep{mgcmqwsl01a, mgcmqwsl01b}.
If all dIs are rotationally supported and dSphs are not \citep[although there
is mounting evidence that this is not strictly true, e.g.,][]
{ddzh04, ggrc06}, then it takes more than just
gas removal to convert a dI to a dSph, since some loss of angular momentum
must also occur. This means that internal mechanisms
of gas removal, such as gas expulsion through galactic winds, or gas exhaustion 
through continued star formation, are not sufficient, and other processes
are needed. Furthermore, one would normally expect to find gas return from
dying stars collecting with time in dSphs, with a return rate thought to be
1\% to 5\% of the stellar mass. Nonetheless, dSphs have extremely low HI 
non-detection limits
which are well below the expected gas mass. This means some other processes
must be at play to remove this gas.

To investigate the morphology-density relation in the Centaurus~A Group,
first the criteria to classify a dwarf as a `transition' dwarf need to be 
clarified. For \citet{sh91} the definition of a
`transition' dwarf was based only on its optical appearances, and any dwarf
exhibiting any characteristics belonging to the other group would be classified
as a `transition' dwarf. This, of course, produced a very
heterogeneous set of objects \citep{ksg99}. \citet{sb84} introduced
a dS0 class, again based purely on visual inspection, but they correctly
inferred that these dwarfs must have a disk component. Indeed, e.g., \citet{aivms05}
showed that dS0s need a Sersic + exponential fit to their 
luminosity profile. In their study of dwarfs in the Virgo cluster \citet{lgwg06} 
creates a new class dEdi to include all dEs with embedded disk features
(hence including the dS0s). However, it seems that disks are so common amongst
dEs that they should be considered a normal feature for an early-type, very much 
like for the bright Elliptical galaxies which often come with 
nuclear disks too \citet{cpf06}, even for systems which are otherwise apparently 'normal'.
Similarly, the presence of HI gas in a dE should not be a sufficient criterion
to classify it as a transition object. \citet{cogw03} find that about 15\%
of Virgo dE are detected in HI, and again this is similar to the detection rate
of big Ellipticals, detected in HI by HIPASS at a rate of 5\%
for Es and 12\% for S0s \citep{s01}. Many `normal' nearby dEs are known to
have some HI (e.g., NGC~205, NGC~185 in the Local Group), although typically dIs
have $\geq 10^{6}$ M$_\odot $ while dEs and dSphs have $< 10^{5}$ M$_\odot $.

Here we will adopt the definition of `transition' dwarf following \citet{m98},
\citet{scm03a}, and \citet{ggh03}, where such a dwarf should have cold gas but no
active star formation, in other words it is detected in HI but not at all 
in H$\alpha $ or with an abnormaly faint flux. 
They represent gas-rich examples of dSphs, and indeed they are found
to fall in the luminosity-metallicity relation near the dSph locus rather 
than the dI one \citep{m98}.  This still is not a
clear-cut way of differentiating these objects as there appears to be 
a continuum of dwarfs from dEs to dIs with various amounts of H$\alpha$ and HI.
Normal dIs, according to the star formation histories constructed from       
their resolved stellar content, are continuously forming stars at a low
level, until they get into a burst episode every few
Gyrs. The presence or absence of just a few HII regions will then determine if
the object at this point of time get classified as a normal dI or a
transition dwarf. The transition dwarfs defined this way will inevitably 
be still a little bit of a mixed collection of objects, with objects who have truly
halted all their star formation activity by depletion of enough gas, and 
those that have still enough gas but are simply in-between such episodes of star
formation. Note though that for nearby dwarfs like in our sample here,
sometimes only one hot star in the galaxy is sufficient to enable its 
detection in H$\alpha$. For example in Pegasus, there are only 2 small HII 
regions detected \citep{sbk97}.  Despite these  
detections it has been classified a transition dwarf by \citep{m98}, because 
the H$\alpha$ flux is extremely faint and its $\tau _{gas}$ is enormous 
(=3220 Gyr). In fact, star forming galaxies follow a trend of increasing
SFRs with decreasing (brighter) magnitudes, equivalent
to a close to constant star formation rate per unit luminosity 
\citep[see Figure 2 of][]{kk07}, but Pegasus, and other transition dwarfs with
detected H$\alpha$ fluxes, falls completely off
this trend found for all others dIs and spirals.

Following these criteria for defining transition dwarfs, there is a total 
of four transition dwarfs in the Centaurus~A Group: ESO~269-G58, UGCA~365, 
ESO~384-G016, and UKS~1424-460.
UGCA~365 has one HII region but its $\tau _{form}$
is enormous ($=$ 1422 Gyr); and UKS~1424-146 is also detected in H$\alpha $ here but
it is only an extended diffuse region, with no obvious ionising source, and
its $\tau _{form}$ is high ($=$ 281 Gyr). Both ESO~269-G58 and ESO~384-G016 are detected
in HI, and have H$\alpha $ detections in the literature which are very low compared
to dwarfs of their magnitude. 
These transition dwarfs add to the five cases in the Sculptor 
Group: SDIG, DDO~6, UGCA~438 \citep[see][]{scm03a}, ESO~294-G10 \citep{jfb98,bou09},
 and ESO~540-G32 \citep{djb08}; 
and six 
in the Local Group: LGS3, Antlia, DDO~210, Pegasus, Phoenix \citep{m98},
 and the recently discovered Leo T \citep{ibe07}. 

Using only the Centaurus~A Group dwarfs with known distances 
\citep[see][]{k07}, the distances to the nearest large galaxy in the 
group for dwarfs of the three types (dI, transition and dE/dSph) are plotted
in Figure~\ref{fig9}. The Centaurus~A Group dwarfs reinforce the trend that 
is seen in the Local Group and Sculptor Group, where dIs are found at a much
larger mean distance from a massive galaxy than the dEs and dSphs. 
Overall, for the three groups, dIs are at a distance of 0.85 $\pm$ 
0.55 Mpc (1 $\sigma$ standard deviation), while dEs and dSphs are at 
0.23 $\pm$ 0.20 Mpc. Interestingly,
the transition dwarfs are found right in the middle at 0.54  $\pm$ 0.31 Mpc.
Kolmogorov-Smirnov tests to determine the significance of the differences    
between the three distributions indicate that indeed all three distributions
are significantly different. The probability that the transition galaxies 
come from the same sample as the dIs is only 3\%. The probability
is even lower, less than 1\%, for the transition galaxies to come from the dEs. 
This is mostly due to the strong peak at $\sim $ 150 kpc in the dEs distributions,
as the vast majority of dEs and dSphs (all but two) 
are satellites of the massive radio galaxy Centaurus~A. The low average distance 
for dEs and dSphs might be biased by the fact that many targeted searches have been
done around bright galaxies, e.g., M31, and therefore will necessarily find 
mostly close-by
dSphs at small angular separation (although only about half of all the Andromeda
dSphs have been included in the Figure since the rest do not yet have good
distance estimates). For the Centaurus~A and Sculptor groups, no such bias 
should apply, as most of the dEs and dSphs were found in all-sky searches 
on POSSII and ESO/SERC plates for nearby dwarf galaxy candidates 
\citep[e.g.,][]{kk98, kk00}, which were not restricted to particular narrow
group region. Moreover, for the transition dwarfs and normal dIs, deep HI scans
were done with HIPASS \citep{b99} over the entire group regions, so the objects
at the periphery of the groups would have had the same chance of being detected
as those nearer the center of the group.

The fact that the mean distance of the transition dwarfs to the nearest
massive galaxy is quite large, 0.54 $\pm$ 0.31 Mpc, poses some difficulties
for most of the proposed scenarios for the transformation of dIs into
dSphs. For example, simulations of gas-dominated dwarf galaxies transforming
 via the combination of ram pressure and tidal stripping
\citep{mmwsm06, mkmw07} successfully produce dSphs with properties compatible 
with observations. However, in this scenario, the original gas-rich dwarf needs
to go through repeated tidal shocks at the pericenter of its orbit, over a
time period of about 10 Gyr, each orbit lasting a few Gyrs, and with 
apocenters significantly smaller than 0.5 Mpc (although in simulations some
occasional extreme satellites on nearly radial orbits can have apocenters 
far exceeding the virial radius of the primary galaxy). It thus seems unlikely
that our transition dwarfs are in the middle of this process, being too far
from their primary galaxy. In fact, many might even be infalling into
the group potential for the first time \citep[as shown for the Local Group 
with orbit tracing by][]{p89}. The crossing times of these groups are several Gyrs,  
and for Sculptor it is even more than half a hubble time at 6.6 Gyr \citep{k05}.

Moreover, simulations
show that at the first tidal shock a strong bar instability is created, which 
funnels the gas into the central few kpcs, producing bursts of star formation.
This is not seen in our transition dwarfs, which do not show more striking bar features 
than normal
dIs. Note that the \citet{mmwsm06, mkmw07} simulations were carried out
on gas-rich dwarfs with NFW halo profiles, while tidal and ram pressure stripping
would be much more effective and quicker on a flat core dwarf \citep[which we would
argue to correspond better to the observations in any case, e.g.,][]{ccf00}.
Similarly, the galaxy harassment scenario is unlikely to have produced
the transition galaxies observed here. Although, in this case, the large distance
of the dwarfs to a main galaxy is not problematic \citep[in the][simulations
the greatest harassment was obtained for galaxies on elongated orbits with 
apocenters typically of 600 kpc]{mkldo96}, the effect of the high-speed close 
encounter is 
clearly visible on the galaxy's morphology after the very first encounter.
This first encounter produces severely disturbed barred spirals, with sharp
and dramatic features, such as tails and rings \citep{mkldo96}. It takes
several encounters before the galaxy shapes itself into a prolate figure, flattened
equally by rotation and random motions. The nearby transition dwarfs do not
show such features. Rather, many exhibit a morphology similar to post-starburst 
objects, with a large
higher surface brightness region where presumably elevated star formation has
last occurred. 

It has been assumed so far that the observed transition dwarfs
might indeed be in some intermediate phase between gas-rich dwarf and dSph.
But one might argue that, since dIs are constantly going on and off with
their star forming activity, that these so-called transition dwarfs might just
be completely normal dIs that happen to be at the extreme tail of low 
SFRs amongst quiescent dIs.
But one would then have to explain why the 'transition' dwarfs are found to have
a smaller distance on average to their primary galaxy compared to normal dIs. 
The natural explanation would be that their quiescent period and SF bursts are 
somehow linked to their positions in 
their orbits, with star formation being mostly triggered at each pericenter passing
(such that the dwarfs that we classify as transitions have just gone through a burst 
and are now in a post-starburst phase with no more active SF).
 But if this were truly the case,
then one would expect to see a similar correlation between the SFRs,
normalised by the area, and the distance to the primary galaxy for {\it all} dIs,
as they would be expected to be affected the same way
  along their orbits. But no trend is found in Figure~\ref{fig8}. 
In view of their smaller average distance to their primary, it
is therefore difficult to see the transition dwarfs as nothing more than
hyperquiescent dwarfs.

The fact that the transition dwarfs are found closer in the groups than
on average other dIs points to an underlying cause related to the
group environment. Something, or some combination of effects, must be
acting on them, with the final outcome being gas depletion and hence
the end of star formation in the galaxy. 
Whatever these effects are (stripping, stirring, harassment etc.),
the first signs of it should be visible on the gas envelope of the
galaxy, the gas being the galaxy's component most vulnerable to
stripping, ejecting, accreting, or funnelling etc. In fact, in the Virgo Cluster, 
Chung et al. (2007) find galaxies with long HI tails 
due to ram pressure stripping
and/or tidal interactions, and it appears that these galaxies begin to lose
their gas already at intermediate distances from M87 (0.6 to 1 Mpc),
similar to the distances of our transition dwarfs to their primary galaxy.  
In a group environment too, as the normal dIs
fall into the group potential perhaps then some signs should start
to show up in their HI envelope as they get closer to the primary galaxies.
One might then think, since the transition dwarfs typically have less gas
than normal dIs \citet{s96}, that perhaps there is a similar trend in normal dIs,
with those at large distances from the group's primary galaxies being 
the most gas-rich and those closer in being gas poorer. Figure~\ref{fig10}
explores this possibility, where the HI mass to light ratio M$_{HI}/L$ is plotted
against the distance of each dwarf from a group's primary galaxy.
There is no such trend visible.
DIs' HI content therefore does not seem  to be depleted
slowly as the dIs approach closer to their primary galaxy, on the way 
to start their transition.  Perhaps what one should be looking for is not
a decrease in total HI mass, but rather a lowering of the HI gas
surface densities, as they are closely linked to star formation thresholds
(see previous section). Any changes in the average HI surface densities
could result in influencing the star formation rate of the dwarf galaxy.

Unfortunately, it is very difficult, with existing data, to study 
quantitatively any major
differences between HI surface densities of transition dwarfs and dIs,
as well as between dIs further or closer to the group's galaxies, because the
few dwarfs which have been imaged in radio with aperture synthesis all have
been observed at different spatial resolutions with different beam sizes.
Hence, a dwarf might have an HI peak just as strong as the others, but if 
it is observed at lower resolution with a larger beam the signal will 
be diluted and the peak HI column density will be underestimated. From the 
sparse data which are available, it does appear that dIs do not show 
dramatic variations in their HI surface densities depending on their 
positions in the groups, although there is a hint that transition dwarfs
do have lower average HI surface densities. For example, for the eight Sculptor 
Group and Centaurus~A Group dwarfs
imaged in HI by \citet{ccf00} with roughly the same beamsizes, the only
transition dwarf in this sample, SDIG, also happens to have the lowest HI surface 
densities. \citet{lsy93} pointed out that LGS3 also has HI column densities 1/10th
those of all the other 9 nearby dIs in their sample. 
There are several transition dwarfs though (e.g., Pegasus, DDO~210 from the 
Local Group), that have very high peak HI column densities, much higher than 
the levels at which there
is star formation in many dIs. In any case, the HI distribution does not seem
to corroborate many of the proposed scenarios for the transition from dI to dSph,
which often involves as a first step the HI being funnelled through a strong bar 
to the centre of the dwarf. 
The HI loss and its consequence the end of star formation must probably happen 
on a very short timescale as the dI falls into the group core, or its orbit 
gets closer to larger galaxies.

\subsection{Comparison with the Local Group and Sculptor Group}

The Centaurus~A Group is now one of the only nearby groups at $<$ 5 Mpc which have
had its dwarf galaxies thoroughly surveyed in H$\alpha$, along with the Sculptor
Group \citep{scm03a}, and the Local Group \citep[see][review]{m98}.
This gives us the chance to compare for the first time the star forming properties
of dwarfs in different group environments. 

Looking through Figures 3 to 10, we can make the following comparisons between
the three groups. Firstly, it appears that the Local Group dwarfs have on average
a lower HI mass to light ratio M$_{HI}/L$, compared to the Centaurus~A Group dwarfs, 
and especially the Sculptor Group dwarfs, which contain several objects 
having M$_{HI}/L$ above 2.  The Local Group lacks these very gas-rich dwarfs. 

Second, from the $SFR/L_B$ from Figure 5 and $SFR/{area}$ of Figure 9, 
it appears that the Local Group is notably different from
the two others in its higher number of dwarf galaxies with high SFRs.
Indeed, the Local Group is rich in dwarfs with presently elevated normalised SFRs, 
such as IC~10 or NGC~6822, which are rarer in other groups. These dwarfs are 
not necessarily classifiable as starbursts, but they nevertheless have SFRs 
much higher than the average 
SFR for a dwarf of the same luminosity and/or HI mass (typically 
$\sim$ 1 $\times$ 10$^{-2}$ $M_{\odot} yr^{-1}$).
One might have expected the Centaurus~A Group to have the largest number of
starbursting and/or high star-forming dwarf galaxies, considering that all the
main members of this group show signs of activity or recent activity. However, 
it appears
that the dwarf population of the group does not follow the trend set by the
larger galaxies. 

The Centaurus~A Group is also the densest of the three groups. One might think that 
its dwarfs may have simply evolved much faster
than the main members, and that their active periods are in the past, with
many dwarfs having already achieved a successful conversion from dI to dSph/dE,
while this process is still more actively taking place in the two other groups.
If this were the case, then detailed star formation histories of dIs and
transition dwarfs should show differences between the Centaurus~A Group and the other
two groups' dwarfs. However, detailed HST CMDs for three dIs of the group 
by \citet{g07} finds that their SFRs have been very low, expect perhaps for 
one object (HIPASSJ1321-31)
which has a peculiar red plume suggesting a `miniburst' about 300 to 500 Myr ago.
In the Local Group, where most of the dIs have high quality CMDs, the clearest 
conclusion
that can be drawn from these CMDs is that no two dIs have the same star formation
history. The five transition dwarfs in particular do not show any recent spike
of star forming activity, except perhaps Antlia in its inner regions only.
If it is truly a burst of star formation which has triggered the process of
transformation in these dwarfs from dI to dSph, then these bursts must have
happened many Gyrs ago, as age indicators from CMDs lose resolution for
ages exceeding a couple of Gyrs. All three Groups show more or less similar numbers
proportionally of transition dwarfs: six for the Local Group, four for the 
Centaurus~A Group, and five for the Sculptor Group.
So the process of transformation of dIs into dSph, or at least into transition 
dwarfs, has the same efficiency in all groups, and the 
timescales to go through this process are probably similar too.

It is surprising to see that it is the Local Group which has
the dwarfs with the highest HI mass to light ratios, 
and the largest number of dwarfs with high 
SFRs. When looking at the basic properties of the groups,
the Local Group has, for example, a crossing time and a density of galaxies 
similar to those of the
Centaurus~A Group, while the Sculptor Group has a considerably longer 
crossing time and lower density.
The Local Group also has a very average total mass in galaxies compared to 
the Centaurus~A Group and the Sculptor Group, as well as an average 
mass to light ratio as a whole \citep{k05}.  
The number of transition dwarfs are all proportionally similar (all this
within small numbers statistics), while one might have expected this number
to scale with 
the crossing times and density of galaxies in the group, since
according to the transition scenarios one needs the dIs to be on closer
orbits with a main galaxy for the various effects to be effective.
The transition dwarf phase therefore seems a widespread phenomena 
in a wide variety of group environment.

\section{Conclusions and Summary}

From our H$\alpha$ imaging survey of dIs in the Centaurus~A Group,
we find the following results:

The Centaurus~A Group dIs do not have enhanced star forming activities,
and do not contain a larger fraction of dwarf starbursts when compared to other
nearby groups such as the Sculptor Group or the Local Group. This is
perhaps surprising given that all large galaxies in the group are
exhibiting or have recently exhibited a period of enhanced star formation.
The gas depletion timescales of the dIs are very large, and except
for a few rare cases, are several times the present age of the universe.

We find that the SFRs of the Centaurus~A Group dIs do not
depend on any global properties of the galaxies, such as magnitude, central 
surface brightness, HI mass-to-light ratio, or B-R color. They do not depend
on local environment either, in particular there is 
no correlation with the distance of the dI to the nearest large galaxy
of the group.

Nonetheless, there is a morphology-density relation in the Centaurus~A Group, 
similarly to the Sculptor Group and Local Group, in the sense that
dEs/dSphs tend to be at small distances from the more massive galaxies
of the group, while dIs are on average at larger distances. Interestingly,
the four transition dwarfs of the Centaurus~A Group (ESO~269-G58, UGCA~365,
ESO~384-G16 and UKS~1424-460) have an average distance intermediate between those
of the dEs/dSphs and dIs. Together with the four transition dwarfs of 
the Sculptor Group and six of the Local Group, they have a quite large
average distance of 0.54$\pm$ 0.31 Mpc. This large distance poses some
difficulty for the most popular scenarios proposed for transforming a dI
into a dE/dSph. Both ram-pressure with tidal stripping, as well as galaxy
harassment, need the dI to be much closer in its orbit to the large galaxy to have
any effect. Also, after the first encounter, they produce transition
objects that appear as strongly barred disturbed objects. If the observed
transition dwarfs are indeed missing links between dIs and dE/dSphs
they are unlikely to have been produced by these mechanisms.

A possible scenario is that the IGM in these groups of galaxies is inhomogeneous,
with regions of higher densities $> 10^{-5}$ cm$^{-3}$, where clumps would be
able to remove the HI from a dI by ram-pressure stripping. 
Several observations point indeed towards such a possibility. According to hydrodynamical 
simulations most low-redshift baryons are expected to be found in a warm-hot 
intergalactic medium (WHIM). This WHIM has already been detected in clusters 
of galaxies, e.g. Takei et al. (2007). In groups of galaxies too,   
OVI absorption
lines studies in the UV \citep{sem03} and X-ray absorption lines studies \citep{ni02}
detect warm temperature ionised gas pervasive to the groups, with
densities possibly as high as 10$^{-4}$ cm$^{-3}$.
This clumpy IGM could be
associated loosely with galaxies (because of past mergers or disturbances,
past bursts etc), or could be simply higher density regions of filaments expected
from the cosmic web \citep{dave2001}. This would also explain the 
existence of dSphs distant from any large galaxies such as Cetus and Tucana.

\acknowledgments

We are very grateful to the referee whose comments have greatly improve
the paper. 
This research has made use of NASA's Astrophysics Data System
Bibliographic Services and the NASA/IPAC Extragalactic Database (NED)
which is operated by the Jet Propulsion Laboratory, California Institute
of Technology, under contract with the National Aeronautics and Space
Administration. 
EDS is grateful for partial support from a NASA LTSARP grant No. NAG5-9221
and the University of Minnesota.
BWM is supported by the Gemini Observatory, which is operated by the 
Association of Universities for Research in Astronomy, Inc., on behalf 
of the international Gemini partnership of Argentina, Australia, Brazil, 
Canada, Chile, the United Kingdom, and the United States of America. 

\appendix
\section{Appendix: Table of HII regions}
The Table~\ref{tbl-2} below lists all the HII regions positions and luminosities
for all the Centaurus~A Group dIs.

{}

\clearpage

\begin{deluxetable}{lcccccc}
\tablenum{1}
\tablewidth{0pt}
\tablecaption{Centaurus A Group Dwarf Irregular Galaxies detected in H$\alpha$
\label{tbl-1}}
\tablehead{
\colhead{Galaxy Name} &  \colhead{R.A. (J2000)} & 
\colhead{Dec.\ (J2000)} & \colhead{V$_{\odot}$} & \colhead{D (Mpc)} & \colhead{Ref.} & \colhead{M(B)}
}
\startdata
ESO~381-G20
            &12:46:00.4 &$-$33:50:17&596 &5.45$\pm 0.44$ & 1 &$-$14.89$\pm 0.05$ \\
DDO~161
             &13:03:17.3 &$-$17:25:20&747 &3.80 &  &$-$15.36$\pm 0.07$ \\
CEN~6
            &13:05:01.0 &$-$40:04:04&619 &5.78$\pm 0.46$ & 1 &$-$12.94$\pm 0.04$ \\
ESO~324-G24      &13:27:35.3 &$-$41:28:50&526 &3.73$\pm 0.43$ & 2 &$-$14.92$\pm 0.04$  \\
UGCA~365     &13:36:30.7 &$-$29:14:11&582 &5.25$\pm 0.43$ & 1 &$-$13.65$\pm 0.09$ \\
ESO~444-G84
           &13:37:20.1 &$-$28:02:46&591 &4.61$\pm 0.46$ & 2 &$-$13.65$\pm 0.04$ \\
NGC~5237      &13:37:38.9 &$-$42:50:51&369 &3.40$\pm 0.27$ & 1 &$-$14.74$\pm 0.04$  \\
IC~4316
         &13:40:18.0 &$-$28:53:40&589 &4.41$\pm 0.44$ & 2 &$-$14.03$\pm 0.04$ \\
NGC~5264
         &13:41:36.9 &$-$29:54:50&487 &4.53$\pm 0.45$ & 2 &$-$16.03$\pm 0.09$ \\
ESO~325-G11 &13:45:00.7 &$-$41:51:32&550 &3.40$\pm 0.39$ & 2 &$-$13.83$\pm 0.07$ \\
ESO~383-G87
        &13:49:18.7 &$-$36:03:41&333 &3.45$\pm 0.27$ & 1 &$-$17.07$\pm 0.09$ \\
NGC~5408
          &14:03:21.4 &$-$41:22:36&506 &4.81$\pm 0.48$ & 2 &$-$16.15$\pm 0.05$ \\
UKS~1424-460 &14:28:03.3 &$-$46:18:13&397 &3.58$\pm 0.33$ & 2 &$-$13.54$\pm 0.11$ \\
ESO~222-G10 &14:35:02.9 &$-$49:25:18&632 &3.80 &&$-$12.85$\pm 0.04$ \\
ESO~272-G25 &14:43:25.5 &$-$44:42:19&624 &3.80 &&$-$13.72$\pm 0.04$ \\
ESO~223-G09 &15:01:01.3 &$-$48:15:51&593 &6.40$\pm 0.51$ & 1 &$-$14.91$\pm 0.06$ \\
ESO~274-G01 &15:14:11.5 &$-$46:47:39&528 &3.05$\pm 0.24$ & 1 &$-$17.25$\pm 0.09$ \\

\enddata
\tablecomments{Heliocentric velocities and magnitudes are from 
\citet{cfcq97}. Distances are from: (1) \citet{k07};
 (2) \citet{k02}.}
\end{deluxetable}

\clearpage

\begin{deluxetable}{lccrc}
\tablenum{2}
\tablecaption{HII Regions Positions and H$\alpha $ Fluxes
\label{tbl-2}}
\tablewidth{0pt}
\tablehead{
\colhead{Galaxy and Number} & \colhead{R.A.} & \colhead{Dec.} &
\colhead{H$\alpha $ Flux} & \colhead{log$L(H\alpha )$} \\
\colhead{} & \multicolumn{2}{c}{(J2000)} &
\colhead{ 10$^{-15}$ ergs cm$^{-2}$ s$^{-1}$} & \colhead{ergs s$^{-1}$}}
\startdata
ESO 381-G020 \#1 & 12:45:53.30 & -33:48:48.5 & 5.2 $\pm$ 1.1 & 37.3 \\
ESO 381-G020 \#2 & 12:45:53.70 & -33:48:52.8 & 8.4 $\pm$ 1.2 & 37.5 \\
ESO 381-G020 \#3 & 12:45:57.40 & -33:49:26.3 & 6.2 $\pm$ 0.8 & 37.3 \\
ESO 381-G020 \#4 & 12:45:58.90 & -33:49:58.3 & 24.0 $\pm$ 1.4 & 37.9 \\
ESO 381-G020 \#5 & 12:45:59.50 & -33:50:01.2 & 5.8 $\pm$ 1.1 & 37.3 \\
ESO 381-G020 \#6 & 12:45:59.00 & -33:49:53.0 & 2.8 $\pm$ 0.6 & 37.0 \\
ESO 381-G020 \#7 & 12:45:59.10 & -33:50:34.1 & 12.2 $\pm$ 1.1 & 37.6 \\
ESO 381-G020 \#8 & 12:45:59.10 & -33:50:37.4 & 9.8 $\pm$ 1.0 & 37.5 \\
ESO 381-G020 \#9 & 12:46:01.50 & -33:50:27.9 & 11.4 $\pm$ 1.1 & 37.6 \\
ESO 381-G020 \#10 & 12:46:01.50 & -33:50:24.3 & 18.8 $\pm$ 1.3 & 37.8 \\
ESO 381-G020 \#11 & 12:46:01.10 & -33:50:31.3 & 2.6 $\pm$ 0.5 & 37.0 \\
ESO 381-G020 \#12 & 12:46:00.90 & -33:50:27.6 & 5.6 $\pm$ 0.7 & 37.3 \\
ESO 381-G020 \#13 & 12:46:01.30 & -33:50:34.6 & 0.7 $\pm$ 0.1 & 36.4 \\
ESO 381-G020 \#14 & 12:46:04.30 & -33:50:35.6 & 4.5 $\pm$ 1.1 & 37.2 \\
ESO 381-G020 \#15 & 12:45:57.60 & -33:49:41.2 & 3.9 $\pm$ 1.2 & 37.1 \\
 & &  & & \\
DDO 161 \#1 & 13:03:02.93 & -17:23:35.1 & 14.7 $\pm$ 3.0 & 37.4 \\
DDO 161 \#2 & 13:03:04.56 & -17:24:01.4 & 1.6 $\pm$ 0.5 & 36.4 \\
DDO 161 \#3 & 13:03:05.89 & -17:24:11.3 & 3.5 $\pm$ 0.4 & 36.8 \\
DDO 161 \#4 & 13:03:06.82 & -17:24:08.7 & 12.7 $\pm$ 1.6 & 37.3 \\
DDO 161 \#5 & 13:03:12.32 & -17:24:43.8 & 6.6 $\pm$ 1.7 & 37.1 \\
DDO 161 \#6 & 13:03:12.62 & -17:25:02.0 & 6.3 $\pm$ 1.7 & 37.0 \\
DDO 161 \#7 & 13:03:14.46 & -17:25:04.3 & 26.3 $\pm$ 1.7 & 37.7 \\
DDO 161 \#8 & 13:03:15.16 & -17:25:07.5 & 82.0 $\pm$ 1.7 & 38.2 \\
DDO 161 \#9 & 13:03:15.50 & -17:24:50.9 & 52.0 $\pm$ 1.7 & 38.0 \\
DDO 161 \#10 & 13:03:14.52 & -17:24:52.7 & 25.1 $\pm$ 1.7 & 37.6 \\
DDO 161 \#11 & 13:03:15.21 & -17:24:59.2 & 5.2 $\pm$ 1.7 & 37.0 \\
DDO 161 \#12 & 13:03:16.13 & -17:24:56.7 & 19.1 $\pm$ 1.7 & 37.5 \\
DDO 161 \#13 & 13:03:16.37 & -17:25:02.4 & 15.3 $\pm$ 1.6 & 37.4 \\
DDO 161 \#14 & 13:03:15.39 & -17:25:16.6 & 5.8 $\pm$ 1.1 & 37.0 \\
DDO 161 \#15 & 13:03:14.51 & -17:24:41.9 & 0.8 $\pm$ 0.1 & 36.1 \\
DDO 161 \#16 & 13:03:14.22 & -17:24:45.2 & 0.8 $\pm$ 0.0 & 36.2 \\
DDO 161 \#17 & 13:03:13.82 & -17:24:52.8 & 1.4 $\pm$ 0.5 & 36.4 \\
DDO 161 \#18 & 13:03:16.60 & -17:25:08.2 & 17.4 $\pm$ 1.6 & 37.5 \\
DDO 161 \#19 & 13:03:17.06 & -17:25:05.7 & 38.6 $\pm$ 1.3 & 37.8 \\
DDO 161 \#20 & 13:03:17.52 & -17:25:06.5 & 22.9 $\pm$ 0.6 & 37.6 \\
DDO 161 \#21 & 13:03:17.24 & -17:25:14.0 & 110.8 $\pm$ 1.7 & 38.3 \\
DDO 161 \#22 & 13:03:17.76 & -17:25:15.6 & 31.9 $\pm$ 1.6 & 37.7 \\
DDO 161 \#23 & 13:03:17.82 & -17:25:23.9 & 25.2 $\pm$ 1.7 & 37.6 \\
DDO 161 \#24 & 13:03:18.51 & -17:25:23.0 & 34.6 $\pm$ 1.7 & 37.8 \\
DDO 161 \#25 & 13:03:18.92 & -17:25:18.8 & 21.3 $\pm$ 1.7 & 37.6 \\
DDO 161 \#26 & 13:03:18.98 & -17:25:25.4 & 25.8 $\pm$ 1.7 & 37.7 \\
DDO 161 \#27 & 13:03:19.04 & -17:25:32.9 & 4.1 $\pm$ 0.5 & 36.9 \\
DDO 161 \#28 & 13:03:19.85 & -17:25:36.9 & 3.7 $\pm$ 0.4 & 36.8 \\
DDO 161 \#29 & 13:03:20.25 & -17:25:37.7 & 7.4 $\pm$ 0.5 & 37.1 \\
DDO 161 \#30 & 13:03:16.03 & -17:25:18.2 & 10.4 $\pm$ 1.7 & 37.3 \\
DDO 161 \#31 & 13:03:15.98 & -17:25:29.9 & 51.6 $\pm$ 1.6 & 38.0 \\
DDO 161 \#32 & 13:03:15.52 & -17:25:33.2 & 13.8 $\pm$ 0.6 & 37.4 \\
DDO 161 \#33 & 13:03:15.23 & -17:25:29.1 & 2.6 $\pm$ 0.5 & 36.7 \\
DDO 161 \#34 & 13:03:16.04 & -17:25:39.8 & 4.5 $\pm$ 1.7 & 36.9 \\
 & &  & & \\
Centaurus 6 \#1 & 13:05:00.24 & -40:04:58.5 & 6.4 $\pm$ 0.8 & 37.4 \\
Centaurus 6 \#2 & 13:05:00.27 & -40:05:02.9 & 3.6 $\pm$ 0.8 & 37.2 \\
Centaurus 6 \#3 & 13:04:59.99 & -40:05:02.0 & 2.1 $\pm$ 0.9 & 36.9 \\
Centaurus 6 \#4 & 13:04:59.07 & -40:05:05.2 & 1.0 $\pm$ 0.7 & 36.6 \\
 & &  & & \\
ESO 324-G024 \#1 & 13:27:36.00 & -41:27:55.0 & 39.7 $\pm$ 1.3 & 37.8 \\
ESO 324-G024 \#2 & 13:27:36.90 & -41:28:12.8 & 29.6 $\pm$ 2.4 & 37.7 \\
ESO 324-G024 \#3 & 13:27:38.40 & -41:29:08.2 & 5.9 $\pm$ 0.3 & 37.0 \\
ESO 324-G024 \#4 & 13:27:38.70 & -41:29:08.2 & 3.8 $\pm$ 0.2 & 36.8 \\
ESO 324-G024 \#5 & 13:27:38.60 & -41:29:57.1 & 7.9 $\pm$ 0.3 & 37.1 \\
 & &  & & \\
UGCA 365 \#1 & 13:36:32.29 & -29:14:18.1 & 1.2 $\pm$ 0.3 & 36.6 \\
 & &  & & \\
ESO 444-G84 \#1 & 13:37:16.63 & -28:02:20.8 & 10.1 $\pm$ 0.7 & 37.4 \\
ESO 444-G84 \#2 & 13:37:18.14 & -28:02:28.9 & 8.0 $\pm$ 0.5 & 37.3 \\
ESO 444-G84 \#3 & 13:37:18.03 & -28:02:31.2 & 3.2 $\pm$ 0.2 & 36.9 \\
ESO 444-G84 \#4 & 13:37:18.40 & -28:02:55.7 & 8.2 $\pm$ 0.6 & 37.3 \\
ESO 444-G84 \#5 & 13:37:18.68 & -28:02:58.2 & 11.7 $\pm$ 0.7 & 37.5 \\
 & &  & & \\
NGC 5237 \#1 & 13:37:38.50 & -42:50:34.5 & 237.9 $\pm$ 16.4 & 38.5 \\
NGC 5237 \#2 & 13:37:37.90 & -42:50:37.7 & 80.2 $\pm$ 5.0 & 38.0 \\
NGC 5237 \#3 & 13:37:37.30 & -42:50:38.6 & 18.7 $\pm$ 2.8 & 37.4 \\
NGC 5237 \#4 & 13:37:37.50 & -42:50:46.8 & 18.4 $\pm$ 3.0 & 37.4 \\
NGC 5237 \#5 & 13:37:38.20 & -42:50:47.0 & 41.2 $\pm$ 7.0 & 37.8 \\
NGC 5237 \#6 & 13:37:39.00 & -42:50:49.1 & 24.3 $\pm$ 6.0 & 37.5 \\
 & &  & & \\
IC 4316 \#1 & 13:40:19.04 & -28:53:10.4 & 12.8 $\pm$ 0.8 & 37.5 \\
IC 4316 \#2 & 13:40:18.97 & -28:53:15.9 & 50.2 $\pm$ 1.1 & 38.1 \\
IC 4316 \#3 & 13:40:18.55 & -28:53:22.9 & 18.6 $\pm$ 1.1 & 37.6 \\
IC 4316 \#4 & 13:40:18.50 & -28:53:26.0 & 21.4 $\pm$ 1.0 & 37.7 \\
IC 4316 \#5 & 13:40:19.26 & -28:53:23.1 & 12.8 $\pm$ 1.0 & 37.5 \\
 & &  & & \\
NGC 5264 \#1 & 13:41:39.33 & -29:54:18.1 & 1.4 $\pm$ 0.4 & 36.5 \\
NGC 5264 \#2 & 13:41:38.13 & -29:54:32.4 & 21.6 $\pm$ 0.9 & 37.7 \\
NGC 5264 \#3 & 13:41:37.78 & -29:54:35.0 & 33.7 $\pm$ 0.8 & 37.9 \\
NGC 5264 \#4 & 13:41:33.99 & -29:55:15.2 & 8.6 $\pm$ 0.8 & 37.3 \\
NGC 5264 \#5 & 13:41:33.86 & -29:55:08.5 & 96.6 $\pm$ 1.9 & 38.4 \\
NGC 5264 \#6 & 13:41:33.73 & -29:55:02.7 & 6.7 $\pm$ 0.4 & 37.2 \\
NGC 5264 \#7 & 13:41:36.43 & -29:54:57.6 & 10.2 $\pm$ 1.1 & 37.4 \\
NGC 5264 \#8 & 13:41:37.17 & -29:54:57.6 & 11.1 $\pm$ 1.1 & 37.4 \\
NGC 5264 \#9 & 13:41:37.52 & -29:54:56.3 & 6.2 $\pm$ 0.4 & 37.2 \\
NGC 5264 \#10 & 13:41:37.76 & -29:54:52.4 & 13.8 $\pm$ 0.9 & 37.5 \\
NGC 5264 \#11 & 13:41:38.26 & -29:54:54.3 & 34.5 $\pm$ 0.8 & 37.9 \\
NGC 5264 \#12 & 13:41:38.70 & -29:54:48.8 & 15.1 $\pm$ 0.9 & 37.6 \\
NGC 5264 \#13 & 13:41:38.31 & -29:54:45.9 & 8.6 $\pm$ 0.8 & 37.3 \\
NGC 5264 \#14 & 13:41:39.02 & -29:54:40.4 & 8.4 $\pm$ 0.8 & 37.3 \\
NGC 5264 \#15 & 13:41:38.72 & -29:54:38.8 & 3.4 $\pm$ 0.5 & 36.9 \\
NGC 5264 \#16 & 13:41:36.69 & -29:55:23.4 & 4.4 $\pm$ 0.4 & 37.0 \\
NGC 5264 \#17 & 13:41:35.48 & -29:54:32.0 & 3.2 $\pm$ 0.4 & 36.9 \\
 & &  & & \\
ESO 325-G011 \#1 & 13:44:56.79 & -41:51:02.5 & 25.4 $\pm$ 3.8 & 37.5 \\
ESO 325-G011 \#2 & 13:44:58.43 & -41:51:07.2 & 75.9 $\pm$ 4.9 & 38.0 \\
ESO 325-G011 \#3 & 13:44:59.71 & -41:51:14.2 & 27.1 $\pm$ 3.8 & 37.6 \\
ESO 325-G011 \#4 & 13:44:58.30 & -41:51:26.3 & 10.4 $\pm$ 3.6 & 37.2 \\
ESO 325-G011 \#5 & 13:45:01.59 & -41:52:04.1 & 34.2 $\pm$ 3.8 & 37.7 \\
ESO 325-G011 \#6 & 13:45:01.67 & -41:52:20.0 & 20.2 $\pm$ 1.8 & 37.4 \\
ESO 325-G011 \#7 & 13:45:06.37 & -41:52:21.2 & 14.2 $\pm$ 3.8 & 37.3 \\
ESO 325-G011 \#8 & 13:45:07.88 & -41:52:30.6 & 18.7 $\pm$ 3.7 & 37.4 \\
 & &  & & \\
ESO 383-G087 \#1 & 13:49:16.49 & -36:02:49.3 & 9.6 $\pm$ 1.3 & 37.1 \\
ESO 383-G087 \#2 & 13:49:15.98 & -36:02:58.4 & 26.4 $\pm$ 1.2 & 37.6 \\
ESO 383-G087 \#3 & 13:49:14.32 & -36:03:04.4 & 3.9 $\pm$ 0.6 & 36.7 \\
ESO 383-G087 \#4 & 13:49:14.39 & -36:03:12.6 & 8.3 $\pm$ 0.5 & 37.1 \\
ESO 383-G087 \#5 & 13:49:15.38 & -36:03:17.9 & 4.3 $\pm$ 0.5 & 36.8 \\
ESO 383-G087 \#6 & 13:49:15.76 & -36:03:19.9 & 11.4 $\pm$ 0.5 & 37.2 \\
ESO 383-G087 \#7 & 13:49:16.17 & -36:03:19.5 & 8.2 $\pm$ 1.2 & 37.1 \\
ESO 383-G087 \#8 & 13:49:18.98 & -36:02:54.0 & 10.2 $\pm$ 1.3 & 37.2 \\
ESO 383-G087 \#9 & 13:49:19.15 & -36:03:00.2 & 9.1 $\pm$ 0.5 & 37.1 \\
ESO 383-G087 \#10 & 13:49:19.50 & -36:03:04.3 & 5.6 $\pm$ 0.5 & 36.9 \\
ESO 383-G087 \#11 & 13:49:18.27 & -36:03:11.5 & 144.3 $\pm$ 3.0 & 38.3 \\
ESO 383-G087 \#12 & 13:49:18.58 & -36:03:04.8 & 13.1 $\pm$ 1.3 & 37.3 \\
ESO 383-G087 \#13 & 13:49:19.03 & -36:03:15.5 & 18.0 $\pm$ 1.5 & 37.4 \\
ESO 383-G087 \#14 & 13:49:18.21 & -36:03:21.8 & 20.0 $\pm$ 1.7 & 37.5 \\
ESO 383-G087 \#15 & 13:49:17.29 & -36:03:13.6 & 28.0 $\pm$ 3.6 & 37.6 \\
ESO 383-G087 \#16 & 13:49:18.66 & -36:03:27.6 & 17.3 $\pm$ 1.9 & 37.4 \\
ESO 383-G087 \#17 & 13:49:17.74 & -36:03:37.1 & 68.9 $\pm$ 2.4 & 38.0 \\
ESO 383-G087 \#18 & 13:49:14.17 & -36:03:42.8 & 20.0 $\pm$ 1.8 & 37.5 \\
ESO 383-G087 \#19 & 13:49:14.35 & -36:04:03.0 & 9.7 $\pm$ 1.4 & 37.1 \\
ESO 383-G087 \#20 & 13:49:14.36 & -36:04:13.8 & 15.5 $\pm$ 2.1 & 37.4 \\
ESO 383-G087 \#21 & 13:49:15.55 & -36:04:08.3 & 37.9 $\pm$ 1.3 & 37.7 \\
ESO 383-G087 \#22 & 13:49:16.02 & -36:04:02.1 & 99.5 $\pm$ 2.4 & 38.2 \\
ESO 383-G087 \#23 & 13:49:15.13 & -36:03:59.7 & 20.1 $\pm$ 1.9 & 37.5 \\
ESO 383-G087 \#24 & 13:49:16.25 & -36:03:53.4 & 19.1 $\pm$ 1.5 & 37.4 \\
ESO 383-G087 \#25 & 13:49:19.98 & -36:04:07.5 & 14.8 $\pm$ 1.9 & 37.3 \\
ESO 383-G087 \#26 & 13:49:22.20 & -36:03:16.1 & 13.4 $\pm$ 1.5 & 37.3 \\
ESO 383-G087 \#27 & 13:49:21.51 & -36:03:01.3 & 13.7 $\pm$ 1.7 & 37.3 \\
ESO 383-G087 \#28 & 13:49:24.46 & -36:02:51.9 & 8.3 $\pm$ 0.5 & 37.1 \\
ESO 383-G087 \#29 & 15:14:07.70 & -46:49:49.7 & 9.2 $\pm$ 0.5 & 37.1 \\
 & &  & & \\
NGC 5408 \#1 & 14:03:28.00 & -41:21:54.6 & 4.0 $\pm$ 0.2 & 37.0 \\
NGC 5408 \#2 & 14:03:27.53 & -41:21:52.7 & 5.3 $\pm$ 0.2 & 37.2 \\
NGC 5408 \#3 & 14:03:26.70 & -41:22:04.1 & 30.4 $\pm$ 0.7 & 37.9 \\
NGC 5408 \#4 & 14:03:26.84 & -41:22:21.8 & 9.6 $\pm$ 0.4 & 37.4 \\
NGC 5408 \#5 & 14:03:26.45 & -41:22:29.3 & 24.6 $\pm$ 2.4 & 37.8 \\
NGC 5408 \#6 & 14:03:25.69 & -41:22:32.1 & 2.6 $\pm$ 0.2 & 36.9 \\
NGC 5408 \#7 & 14:03:26.01 & -41:22:34.1 & 9.2 $\pm$ 0.4 & 37.4 \\
NGC 5408 \#8 & 14:03:26.31 & -41:22:39.8 & 49.4 $\pm$ 0.4 & 38.1 \\
NGC 5408 \#9 & 14:03:25.94 & -41:22:40.1 & 13.3 $\pm$ 0.4 & 37.6 \\
NGC 5408 \#10 & 14:03:27.24 & -41:22:48.0 & 5.5 $\pm$ 0.2 & 37.2 \\
NGC 5408 \#11 & 14:03:25.11 & -41:21:55.4 & 16.0 $\pm$ 1.6 & 37.6 \\
NGC 5408 \#12 & 14:03:24.05 & -41:21:55.8 & 21.5 $\pm$ 0.6 & 37.8 \\
NGC 5408 \#13 & 14:03:23.66 & -41:21:53.6 & 4.3 $\pm$ 0.2 & 37.1 \\
NGC 5408 \#14 & 14:03:23.46 & -41:21:52.8 & 5.7 $\pm$ 0.2 & 37.2 \\
NGC 5408 \#15 & 14:03:23.10 & -41:21:55.3 & 4.2 $\pm$ 0.2 & 37.1 \\
NGC 5408 \#16 & 14:03:22.81 & -41:22:03.1 & 61.5 $\pm$ 1.3 & 38.2 \\
NGC 5408 \#17 & 14:03:22.29 & -41:22:03.4 & 12.8 $\pm$ 0.2 & 37.6 \\
NGC 5408 \#18 & 14:03:23.70 & -41:22:25.7 & 13.6 $\pm$ 0.6 & 37.6 \\
NGC 5408 \#19 & 14:03:23.21 & -41:22:30.7 & 7.2 $\pm$ 0.4 & 37.3 \\
NGC 5408 \#20 & 14:03:23.31 & -41:22:34.0 & 4.5 $\pm$ 0.2 & 37.1 \\
NGC 5408 \#21 & 14:03:24.22 & -41:22:35.3 & 2.3 $\pm$ 0.1 & 36.8 \\
NGC 5408 \#22 & 14:03:23.95 & -41:22:38.1 & 3.3 $\pm$ 0.1 & 37.0 \\
NGC 5408 \#23 & 14:03:23.96 & -41:22:44.5 & 5.0 $\pm$ 0.3 & 37.1 \\
NGC 5408 \#24 & 14:03:23.56 & -41:22:39.0 & 36.7 $\pm$ 0.3 & 38.0 \\
NGC 5408 \#25 & 14:03:23.02 & -41:22:38.7 & 21.5 $\pm$ 0.4 & 37.8 \\
NGC 5408 \#26 & 14:03:23.32 & -41:22:41.8 & 7.6 $\pm$ 0.1 & 37.3 \\
NGC 5408 \#27 & 14:03:23.42 & -41:22:47.8 & 17.9 $\pm$ 0.7 & 37.7 \\
NGC 5408 \#28 & 14:03:22.68 & -41:22:35.4 & 8.3 $\pm$ 0.2 & 37.4 \\
NGC 5408 \#29 & 14:03:22.51 & -41:22:38.5 & 26.0 $\pm$ 0.2 & 37.9 \\
NGC 5408 \#30 & 14:03:22.83 & -41:22:44.6 & 6.7 $\pm$ 0.2 & 37.3 \\
NGC 5408 \#31 & 14:03:22.49 & -41:22:49.3 & 14.6 $\pm$ 0.3 & 37.6 \\
NGC 5408 \#32 & 14:03:22.24 & -41:22:44.9 & 18.5 $\pm$ 0.2 & 37.7 \\
NGC 5408 \#33 & 14:03:21.77 & -41:22:43.5 & 5.5 $\pm$ 0.2 & 37.2 \\
NGC 5408 \#34 & 14:03:21.48 & -41:22:44.9 & 2.7 $\pm$ 0.2 & 36.9 \\
NGC 5408 \#35 & 14:03:22.75 & -41:22:32.4 & 3.9 $\pm$ 0.1 & 37.0 \\
NGC 5408 \#36 & 14:03:21.51 & -41:22:50.5 & 17.4 $\pm$ 0.4 & 37.7 \\
NGC 5408 \#37 & 14:03:21.06 & -41:22:36.1 & 59.6 $\pm$ 0.9 & 38.2 \\
NGC 5408 \#38 & 14:03:20.64 & -41:22:40.9 & 47.2 $\pm$ 0.4 & 38.1 \\
NGC 5408 \#39 & 14:03:20.89 & -41:22:45.3 & 19.2 $\pm$ 0.3 & 37.7 \\
NGC 5408 \#40 & 14:03:21.06 & -41:22:50.0 & 42.5 $\pm$ 0.5 & 38.1 \\
NGC 5408 \#41 & 14:03:20.30 & -41:22:46.2 & 20.0 $\pm$ 0.4 & 37.7 \\
NGC 5408 \#42 & 14:03:21.13 & -41:22:42.2 & 12.1 $\pm$ 0.3 & 37.5 \\
NGC 5408 \#43 & 14:03:19.78 & -41:22:28.0 & 37.6 $\pm$ 0.7 & 38.0 \\
NGC 5408 \#44 & 14:03:20.25 & -41:22:38.7 & 8.0 $\pm$ 0.2 & 37.3 \\
NGC 5408 \#45 & 14:03:19.93 & -41:22:40.9 & 30.2 $\pm$ 0.3 & 37.9 \\
NGC 5408 \#46 & 14:03:19.30 & -41:22:46.0 & 483.5 $\pm$ 1.1 & 39.1 \\
NGC 5408 \#47 & 14:03:18.59 & -41:22:52.1 & 1874.1 $\pm$ 12.1 & 39.7 \\
NGC 5408 \#48 & 14:03:19.32 & -41:22:50.9 & 401.0 $\pm$ 0.8 & 39.0 \\
NGC 5408 \#49 & 14:03:19.86 & -41:22:46.7 & 53.1 $\pm$ 0.3 & 38.2 \\
NGC 5408 \#50 & 14:03:17.98 & -41:22:54.1 & 48.8 $\pm$ 0.4 & 38.1 \\
NGC 5408 \#51 & 14:03:19.52 & -41:22:59.8 & 40.2 $\pm$ 0.3 & 38.0 \\
NGC 5408 \#52 & 14:03:17.61 & -41:22:53.6 & 27.0 $\pm$ 0.2 & 37.9 \\
NGC 5408 \#53 & 14:03:17.91 & -41:22:58.0 & 24.8 $\pm$ 0.4 & 37.8 \\
NGC 5408 \#54 & 14:03:18.10 & -41:23:02.1 & 40.8 $\pm$ 0.4 & 38.1 \\
NGC 5408 \#55 & 14:03:17.49 & -41:23:01.6 & 20.4 $\pm$ 0.3 & 37.8 \\
NGC 5408 \#56 & 14:03:17.76 & -41:23:07.7 & 15.5 $\pm$ 0.3 & 37.6 \\
NGC 5408 \#57 & 14:03:17.34 & -41:22:46.7 & 16.1 $\pm$ 0.4 & 37.7 \\
NGC 5408 \#58 & 14:03:20.16 & -41:22:52.5 & 3.9 $\pm$ 0.2 & 37.0 \\
NGC 5408 \#59 & 14:03:20.09 & -41:22:57.2 & 31.3 $\pm$ 0.3 & 37.9 \\
NGC 5408 \#60 & 14:03:20.72 & -41:22:56.3 & 26.6 $\pm$ 0.3 & 37.9 \\
NGC 5408 \#61 & 14:03:18.15 & -41:23:35.5 & 22.5 $\pm$ 0.6 & 37.8 \\
NGC 5408 \#62 & 14:03:18.53 & -41:22:38.8 & 3.0 $\pm$ 0.2 & 36.9 \\
NGC 5408 \#63 & 14:03:17.61 & -41:22:37.8 & 5.6 $\pm$ 0.1 & 37.2 \\
 & &  & &  \\
UKS 1424-460 \#1 & 14:28:02.00 & -46:17:57.0 & 11.9 $\pm$ 1.0 & 37.3 \\
 & &  & & \\
ESO 222-G010 \#1 & 14:35:01.80 & -49:25:21.7 & 13.6 $\pm$ 1.1 & 37.4 \\
ESO 222-G010 \#2 & 14:35:02.20 & -49:25:24.8 & 10.0 $\pm$ 0.9 & 37.2 \\
ESO 222-G010 \#3 & 14:35:02.76 & -49:25:24.4 & 11.3 $\pm$ 1.0 & 37.3 \\
ESO 222-G010 \#4 & 14:35:02.58 & -49:25:28.4 & 46.6 $\pm$ 0.5 & 37.9 \\
ESO 222-G010 \#5 & 14:35:02.73 & -49:25:15.6 & 47.7 $\pm$ 2.5 & 37.9 \\
 & &  & &  \\
ESO 272-G025 \#1 & 14:43:24.10 & -44:42:25.4 & 8.3 $\pm$ 1.2 & 37.2 \\
ESO 272-G025 \#2 & 14:43:24.45 & -44:42:26.1 & 9.3 $\pm$ 1.5 & 37.2 \\
ESO 272-G025 \#3 & 14:43:24.30 & -44:42:19.9 & 4.8 $\pm$ 1.0 & 36.9 \\
ESO 272-G025 \#4 & 14:43:25.08 & -44:42:24.1 & 15.4 $\pm$ 3.2 & 37.4 \\
ESO 272-G025 \#5 & 14:43:24.98 & -44:42:16.1 & 21.9 $\pm$ 1.6 & 37.6 \\
ESO 272-G025 \#6 & 14:43:26.29 & -44:42:09.4 & 17.9 $\pm$ 1.7 & 37.5 \\
ESO 272-G025 \#7 & 14:43:25.96 & -44:42:12.0 & 3.6 $\pm$ 0.7 & 36.8 \\
ESO 272-G025 \#8 & 14:43:26.34 & -44:42:13.9 & 2.6 $\pm$ 0.5 & 36.7 \\
ESO 272-G025 \#9 & 14:43:26.42 & -44:42:02.7 & 19.2 $\pm$ 2.5 & 37.5 \\
ESO 272-G025 \#10 & 14:43:24.53 & -44:42:14.1 & 2.6 $\pm$ 0.7 & 36.7 \\
 & &  & & \\
ESO 223-G009 \#1 & 15:01:17.54 & -48:16:40.0 & 19.3 $\pm$ 5.2 & 38.0 \\
ESO 223-G009 \#2 & 15:01:16.88 & -48:16:39.6 & 7.6 $\pm$ 1.0 & 37.6 \\
ESO 223-G009 \#3 & 15:01:14.83 & -48:19:39.5 & 69.4 $\pm$ 3.3 & 38.5 \\
ESO 223-G009 \#4 & 15:01:16.45 & -48:19:34.3 & 6.4 $\pm$ 1.1 & 37.5 \\
ESO 223-G009 \#5 & 15:01:16.08 & -48:19:31.4 & 18.1 $\pm$ 0.7 & 38.0 \\
ESO 223-G009 \#6 & 15:01:16.70 & -48:19:26.6 & 2.7 $\pm$ 1.2 & 37.1 \\
ESO 223-G009 \#7 & 15:01:15.89 & -48:19:23.3 & 6.3 $\pm$ 1.4 & 37.5 \\
ESO 223-G009 \#8 & 15:01:11.76 & -48:18:13.7 & 7.7 $\pm$ 1.0 & 37.6 \\
ESO 223-G009 \#9 & 15:01:04.33 & -48:17:55.1 & 199.3 $\pm$ 1.2 & 39.0 \\
ESO 223-G009 \#10 & 15:01:04.66 & -48:17:50.0 & 55.5 $\pm$ 0.7 & 38.4 \\
ESO 223-G009 \#11 & 15:01:03.12 & -48:17:49.7 & 9.6 $\pm$ 1.5 & 37.7 \\
ESO 223-G009 \#12 & 15:01:04.91 & -48:17:44.4 & 20.9 $\pm$ 1.4 & 38.0 \\
ESO 223-G009 \#13 & 15:01:05.91 & -48:17:55.8 & 36.3 $\pm$ 0.7 & 38.3 \\
ESO 223-G009 \#14 & 15:01:05.36 & -48:17:50.7 & 40.1 $\pm$ 0.7 & 38.3 \\
ESO 223-G009 \#15 & 15:01:15.88 & -48:17:26.8 & 20.6 $\pm$ 1.5 & 38.0 \\
ESO 223-G009 \#16 & 15:01:18.06 & -48:18:33.8 & 5.6 $\pm$ 1.4 & 37.4 \\
ESO 223-G009 \#17 & 15:01:09.81 & -48:17:58.0 & 19.0 $\pm$ 0.7 & 38.0 \\
ESO 223-G009 \#18 & 15:01:11.42 & -48:17:49.1 & 22.0 $\pm$ 0.6 & 38.0 \\
ESO 223-G009 \#19 & 15:01:10.75 & -48:17:48.3 & 18.6 $\pm$ 0.7 & 38.0 \\
ESO 223-G009 \#20 & 15:01:10.09 & -48:17:46.2 & 9.7 $\pm$ 1.4 & 37.7 \\
ESO 223-G009 \#21 & 15:01:09.65 & -48:17:44.7 & 6.1 $\pm$ 1.4 & 37.5 \\
ESO 223-G009 \#22 & 15:01:07.44 & -48:17:33.0 & 15.8 $\pm$ 0.7 & 37.9 \\
ESO 223-G009 \#23 & 15:01:07.14 & -48:17:19.7 & 13.5 $\pm$ 0.7 & 37.8 \\
ESO 223-G009 \#24 & 15:01:07.13 & -48:17:4.2 & 13.0 $\pm$ 1.1 & 37.8 \\
ESO 223-G009 \#25 & 15:01:04.99 & -48:16:46.6 & 9.1 $\pm$ 0.9 & 37.7 \\
ESO 223-G009 \#26 & 15:01:02.80 & -48:17:6.5 & 53.8 $\pm$ 0.6 & 38.4 \\
ESO 223-G009 \#27 & 15:01:02.35 & -48:16:59.9 & 23.8 $\pm$ 0.6 & 38.1 \\
ESO 223-G009 \#28 & 15:01:02.57 & -48:16:51.1 & 26.9 $\pm$ 0.5 & 38.1 \\
ESO 223-G009 \#29 & 15:01:03.37 & -48:16:45.2 & 11.7 $\pm$ 0.9 & 37.8 \\
ESO 223-G009 \#30 & 15:01:03.14 & -48:16:34.1 & 19.1 $\pm$ 2.5 & 38.0 \\
ESO 223-G009 \#31 & 15:01:06.05 & -48:15:47.6 & 10.5 $\pm$ 1.6 & 37.7 \\
ESO 223-G009 \#32 & 15:01:05.30 & -48:17:9.4 & 3.3 $\pm$ 1.3 & 37.2 \\
ESO 223-G009 \#33 & 15:01:04.85 & -48:17:4.3 & 5.0 $\pm$ 1.3 & 37.4 \\
ESO 223-G009 \#34 & 15:01:03.75 & -48:16:59.9 & 13.3 $\pm$ 0.8 & 37.8 \\
ESO 223-G009 \#35 & 15:01:07.52 & -48:17:45.5 & 27.4 $\pm$ 0.5 & 38.1 \\
ESO 223-G009 \#36 & 15:01:10.28 & -48:18:44.4 & 14.4 $\pm$ 0.7 & 37.9 \\
ESO 223-G009 \#37 & 15:01:09.43 & -48:17:41.0 & 10.7 $\pm$ 1.5 & 37.7 \\
ESO 223-G009 \#38 & 15:01:03.56 & -48:17:49.3 & 10.3 $\pm$ 1.5 & 37.7 \\
ESO 223-G009 \#39 & 15:01:04.96 & -48:17:59.6 & 9.1 $\pm$ 1.4 & 37.7 \\
ESO 223-G009 \#40 & 15:01:03.46 & -48:17:10.2 & 6.9 $\pm$ 1.4 & 37.5 \\
 & &  & & \\
ESO 274-G001 \#1 & 15:14:30.40 & -46:44:40.7 & 100.0 $\pm$ 4.4 & 38.0 \\
ESO 274-G001 \#2 & 15:14:28.80 & -46:44:56.1 & 39.6 $\pm$ 2.9 & 37.6 \\
ESO 274-G001 \#3 & 15:14:29.20 & -46:45:08.2 & 7.6 $\pm$ 0.7 & 36.9 \\
ESO 274-G001 \#4 & 15:14:28.40 & -46:45:09.9 & 10.6 $\pm$ 1.3 & 37.1 \\
ESO 274-G001 \#5 & 15:14:17.60 & -46:47:16.6 & 20.3 $\pm$ 1.3 & 37.4 \\
ESO 274-G001 \#6 & 15:14:17.70 & -46:47:23.8 & 27.2 $\pm$ 1.2 & 37.5 \\
ESO 274-G001 \#7 & 15:14:16.60 & -46:47:36.8 & 15.9 $\pm$ 1.3 & 37.3 \\
ESO 274-G001 \#8 & 15:14:15.60 & -46:47:57.0 & 35.2 $\pm$ 3.4 & 37.6 \\
ESO 274-G001 \#9 & 15:14:14.60 & -46:48:17.3 & 148.0 $\pm$ 1.9 & 38.2 \\
ESO 274-G001 \#10 & 15:14:14.80 & -46:48:09.2 & 46.0 $\pm$ 1.5 & 37.7 \\
ESO 274-G001 \#11 & 15:14:13.70 & -46:48:12.5 & 89.7 $\pm$ 1.1 & 38.0 \\
ESO 274-G001 \#12 & 15:14:13.20 & -46:48:19.8 & 31.1 $\pm$ 1.3 & 37.5 \\
ESO 274-G001 \#13 & 15:14:14.00 & -46:48:23.0 & 32.3 $\pm$ 0.8 & 37.6 \\
ESO 274-G001 \#14 & 15:14:13.80 & -46:48:35.9 & 27.1 $\pm$ 2.2 & 37.5 \\
ESO 274-G001 \#15 & 15:14:13.00 & -46:48:30.3 & 15.8 $\pm$ 1.1 & 37.2 \\
ESO 274-G001 \#16 & 15:14:13.00 & -46:49:02.6 & 28.7 $\pm$ 2.6 & 37.5 \\
ESO 274-G001 \#17 & 15:14:12.70 & -46:48:58.6 & 14.7 $\pm$ 0.5 & 37.2 \\
ESO 274-G001 \#18 & 15:14:12.00 & -46:48:56.2 & 23.5 $\pm$ 0.7 & 37.4 \\
ESO 274-G001 \#19 & 15:14:11.10 & -46:48:53.8 & 39.5 $\pm$ 1.1 & 37.6 \\
ESO 274-G001 \#20 & 15:14:11.00 & -46:49:05.9 & 32.9 $\pm$ 1.2 & 37.6 \\
ESO 274-G001 \#21 & 15:14:12.00 & -46:49:03.5 & 37.4 $\pm$ 1.2 & 37.6 \\
ESO 274-G001 \#22 & 15:14:12.70 & -46:49:10.7 & 52.5 $\pm$ 1.3 & 37.8 \\
ESO 274-G001 \#23 & 15:14:10.00 & -46:49:02.0 & 446.2 $\pm$ 3.4 & 38.7 \\
ESO 274-G001 \#24 & 15:14:10.50 & -46:49:22.9 & 9.5 $\pm$ 0.6 & 37.0 \\
ESO 274-G001 \#25 & 15:14:09.80 & -46:49:19.7 & 9.8 $\pm$ 0.6 & 37.0 \\
ESO 274-G001 \#26 & 15:14:09.30 & -46:49:17.3 & 11.3 $\pm$ 0.5 & 37.1 \\
ESO 274-G001 \#27 & 15:14:08.80 & -46:49:20.6 & 28.6 $\pm$ 1.6 & 37.5 \\
ESO 274-G001 \#28 & 15:14:08.30 & -46:49:56.1 & 30.7 $\pm$ 1.1 & 37.5 \\
ESO 274-G001 \#29 & 15:14:07.70 & -46:49:49.7 & 26.2 $\pm$ 1.3 & 37.5 \\
ESO 274-G001 \#30 & 15:14:05.20 & -46:50:12.5 & 54.9 $\pm$ 1.5 & 37.8 \\
ESO 274-G001 \#31 & 15:14:05.20 & -46:50:18.9 & 27.9 $\pm$ 0.6 & 37.5 \\
ESO 274-G001 \#32 & 15:14:05.20 & -46:50:26.2 & 16.2 $\pm$ 0.7 & 37.3 \\
ESO 274-G001 \#33 & 15:13:58.90 & -46:51:46.5 & 28.1 $\pm$ 2.4 & 37.5 \\
ESO 274-G001 \#34 & 15:14:14.00 & -46:48:00.4 & 37.9 $\pm$ 3.5 & 37.6 \\
\enddata
\end{deluxetable}

\clearpage

\begin{deluxetable}{lccccccc}
\tablenum{3}
\pagestyle{empty}
\tablecaption{Star Formation Properties of Centaurus A Group dI Galaxies\label{tbl-3}}
\tablewidth{0pt}
\tablehead{
\colhead{Galaxy} & \colhead{SFR} & \colhead{SFR/L(B)} & 
\colhead{$\tau_{form}$} &
\colhead{M(HI)\tablenotemark{a}} & 
\colhead{M(HI)/L(B)} & \colhead{$\tau_{gas}$\tablenotemark{b}} &
\colhead{SFR/area} \\
\colhead{} &   \colhead{M$_{\odot}$ yr$^{-1}$} & 
\colhead{M$_{\odot}$ yr$^{-1}$ L$_{\odot}^{-1}$} &
\colhead{Gyr} & 
\colhead{10$^6$ M$_{\odot}$}  & \colhead{M$_{\odot}$/L$_{\odot}$} & 
\colhead{Gyr} & \colhead{M$_{\odot}$ yr$^{-1}$ pc$^{-2}$}
}
\startdata
ESO~321-G14\tablenotemark{c} & $6.5\times 10^{-4}$ & $4.9\times 10^{-11}$  & 20
            & 15 & 1.13 & 25 &  \\ 
ESO~381-G20 & $3.4\times 10^{-3}$ & $2.5\times 10^{-11}$  & 41
            & 224  & 1.59 & 85 & -9.30 \\ 
UGCA~319\tablenotemark{c} & $1.4\times 10^{-4}$ & $5.9\times 10^{-12}$  & 158 
            & 24.0 & 1.04 & 200 &  \\ 
DDO~161 & $1.0\times 10^{-2}$ & $4.8\times 10^{-11}$  & 21
            & 375  & 1.73 & 48 & -8.45 \\ 
CEN~6        & $4.2\times 10^{-4}$ & $1.8\times 10^{-11}$  & 56
            & 34.7 & 1.48 & 110 & -9.72 \\ 
ESO~269-G58\tablenotemark{d} & $2.4\times 10^{-4}$ & $7.9\times 10^{-13}$  & 1267  
            & 24.8 & 0.08 & 136.4 &  \\ 
AM1321-304\tablenotemark{c} & $4.5\times 10^{-5}$ & $6.2\times 10^{-12}$  & 158 
            & 20.0 & 2.74 & 398 &  \\ 
ESO~324-G24 & $1.9\times 10^{-3}$ & $1.3\times 10^{-11}$  & 77 
            & 171 & 1.18 & 120 & -9.46 \\ 
UGCA~365    & $3.2\times 10^{-5}$ & $7.0\times 10^{-13}$  & 1422 
            & 29.8 & 0.67 & 1250  & -10.83 \\ 
ESO~444-G84 & $8.3\times 10^{-4}$ & $1.9\times 10^{-11}$  & 54
            & 98.1 & 2.18 & 155 & -9.22 \\ 
NGC~5237    & $4.6\times 10^{-3}$ & $3.8\times 10^{-11}$  & 27
            & 20.7  & 0.17 & 5.9 & -7.90 \\ 
IC~4316     & $2.1\times 10^{-3}$ & $3.4\times 10^{-11}$  & 30
            & 35.7 & 0.56 & 22 & -8.77 \\ 
NGC~5264    & $5.9\times 10^{-3}$ & $1.5\times 10^{-11}$  & 68
            & 66.2 & 0.16 & 14.8 & -8.70 \\ 
ESO~325-G11     & $2.5\times 10^{-3}$ & $4.7\times 10^{-11}$ & 21.4
            & 69.2 & 1.30 & 37 & -9.70 \\ 
ESO~383-G87 & $8.4\times 10^{-3}$ & $7.9\times 10^{-12}$  & 126
            & 71.2  & 0.07 & 11.3 & -8.54 \\ 
ESO~384-G16\tablenotemark{e} & $2.2\times 10^{-4}$ & $6.6\times 10^{-12}$  & 152  
            & 6.5 & 0.19 & 39 &  \\ 
NGC~5408 & $8.8\times 10^{-2}$ & $2.0\times 10^{-10}$  & 5.1
            & 357  & 0.80 & 5.3 & -7.35 \\ 
UKS~1424-460 & $1.4\times 10^{-4}$ & $3.6\times 10^{-12}$  & 281
            & 58.4  & 1.44 & 534 & -11.45 \\ 
ESO~222-G10 & $1.8\times 10^{-3}$ & $8.3\times 10^{-11}$  & 12.1
            & 31.0  & 1.44 & 23 & -9.14 \\ 
ESO~272-G25 & $1.5\times 10^{-3}$ & $3.0\times 10^{-11}$  & 33
            & $<$6.5  & $<$0.14 & $<$5.9 & -8.48 \\ 
ESO~223-G09 & $3.4\times 10^{-2}$ & $8.6\times 10^{-11}$  & 11.6
            & 928  & 2.28 & 35 & -8.28 \\ 
ESO~274-G01 & $1.5\times 10^{-2}$ & $1.2\times 10^{-11}$  & 82
            & 256  & 0.21 & 22 & -8.95 \\ 
\enddata

\tablenotetext{a}{Total galaxy HI mass from \citet{cfcq97},
adjusted to the distance in Table 1} 
\tablenotetext{b}{$\tau_{gas}$ is the gas depletion time scale $=$ 
(Total Gas Mass)/(SFR),
where the total gas mass is 1.32 $\times$ M(HI) to account for He}
\tablenotetext{c}{From Bouchard et al 2009}
\tablenotetext{d}{H$\alpha$ data from \citet{ph86}, HI mass from \citet{b99}, using the nominal distance of 3.8 Mpc.}
\tablenotetext{e}{H$\alpha$ data from \citet{bou09}, HI mass from \citet{b06}, using a distance of 4.23 Mpc
from \citet{j00}}.

\end{deluxetable}

\clearpage

\begin{figure}
\vbox to7.2in{\rule{0pt}{2.6in}}
\includegraphics{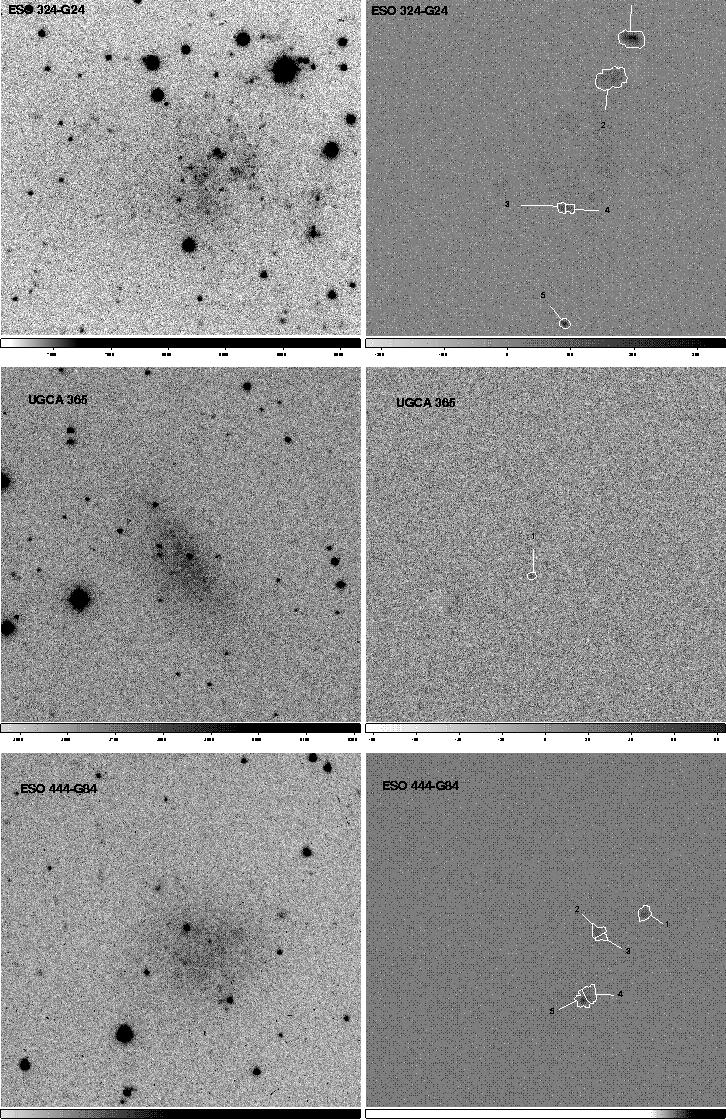}

\end{figure}

\begin{figure}
\vbox to7.2in{\rule{0pt}{2.6in}}
\includegraphics{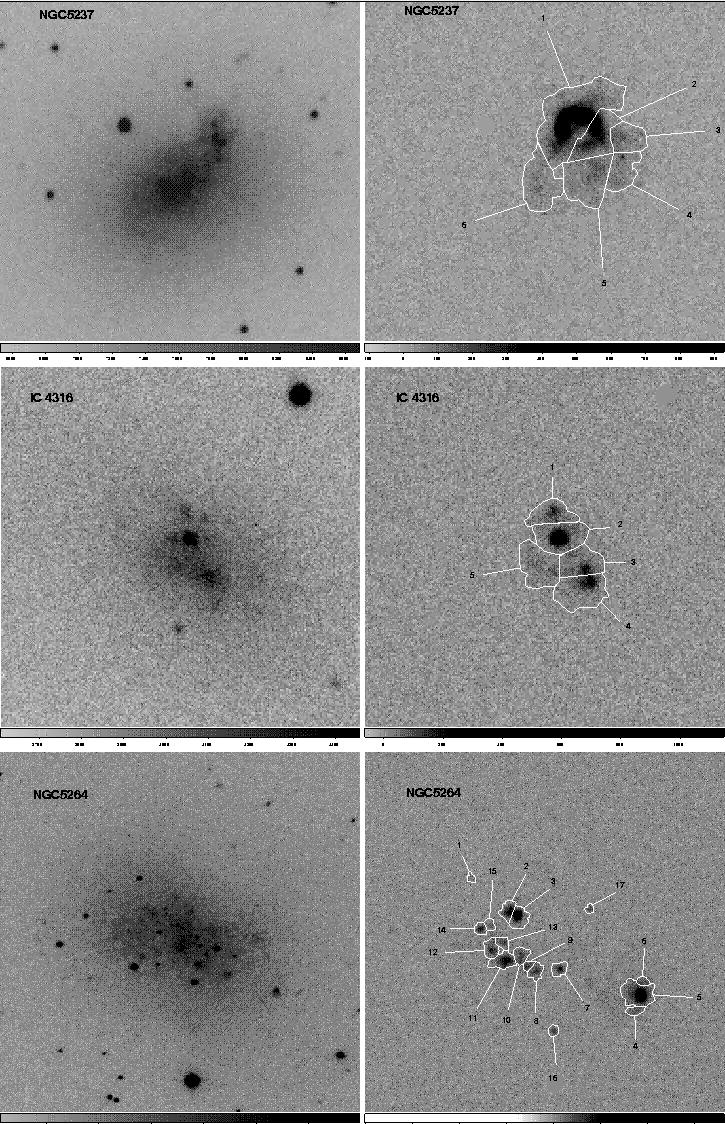}

\end{figure}

\begin{figure}
\vbox to7.2in{\rule{0pt}{2.6in}}
\includegraphics{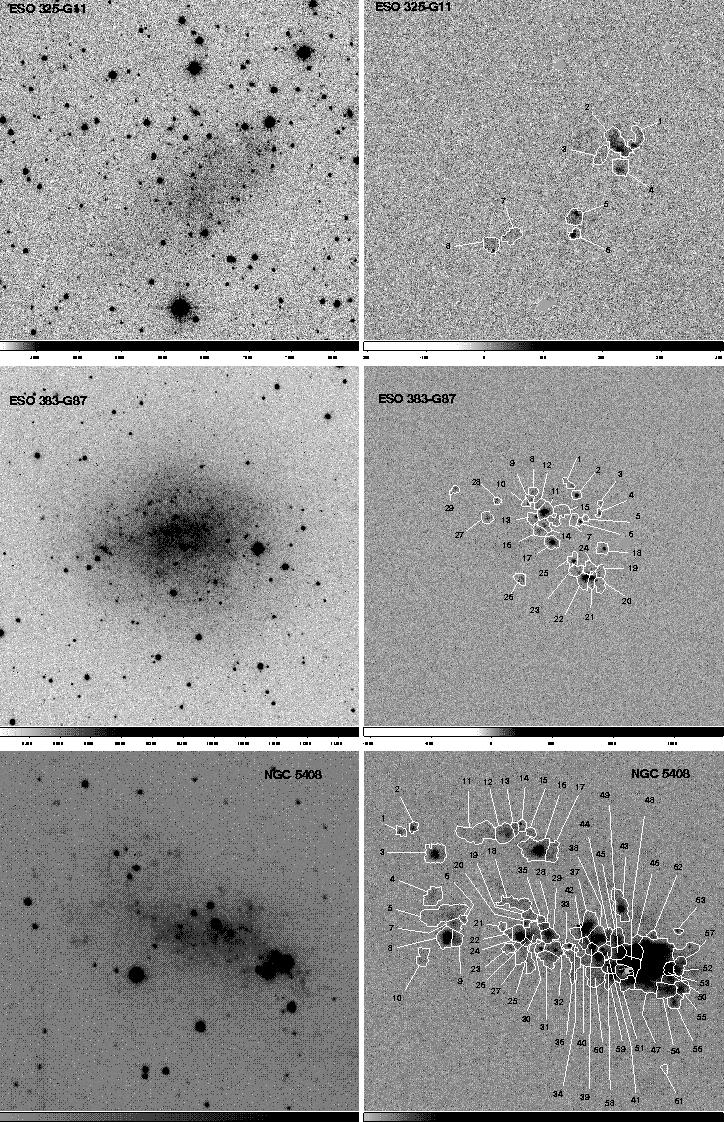}

\end{figure}

\begin{figure}
\vbox to7.2in{\rule{0pt}{2.6in}}
\includegraphics{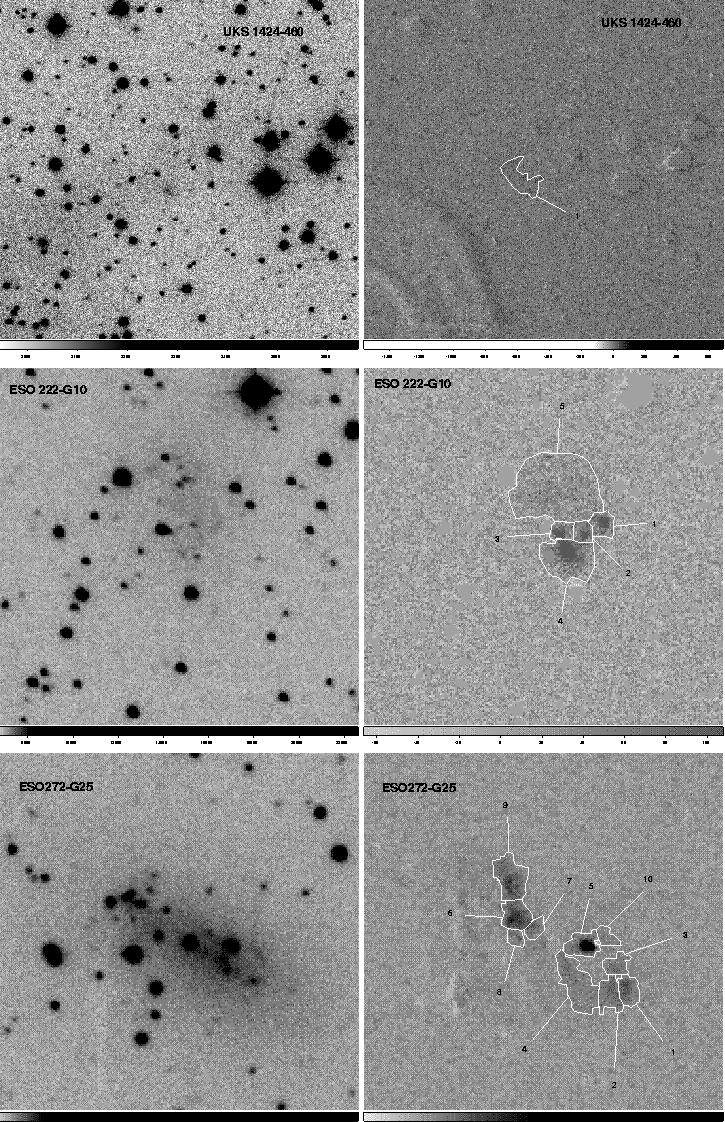}

\end{figure}

\begin{figure}
\vbox to7.2in{\rule{0pt}{2.6in}}
\includegraphics{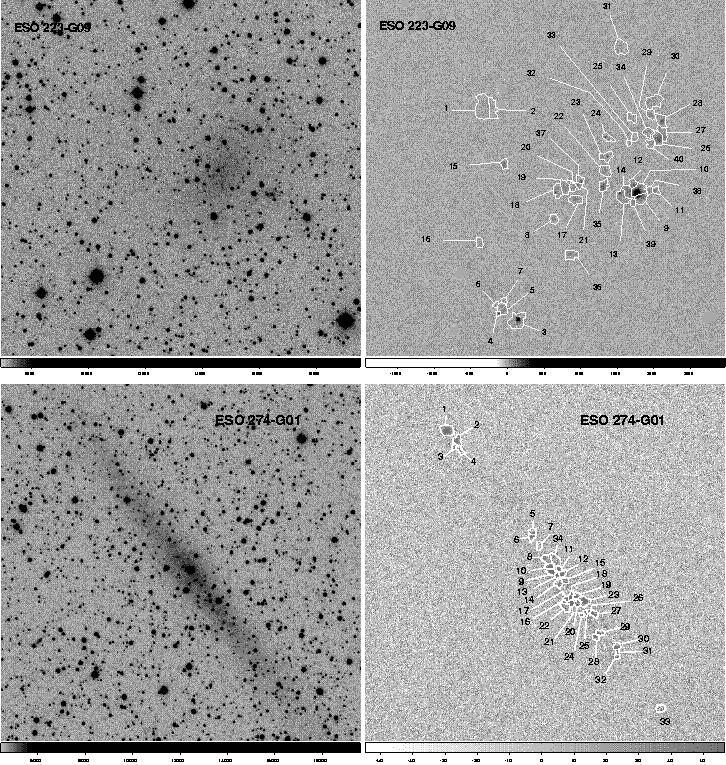}

\end{figure}

\clearpage 

\begin {figure}
\vbox to7.2in{\rule{0pt}{2.6in}}
\includegraphics{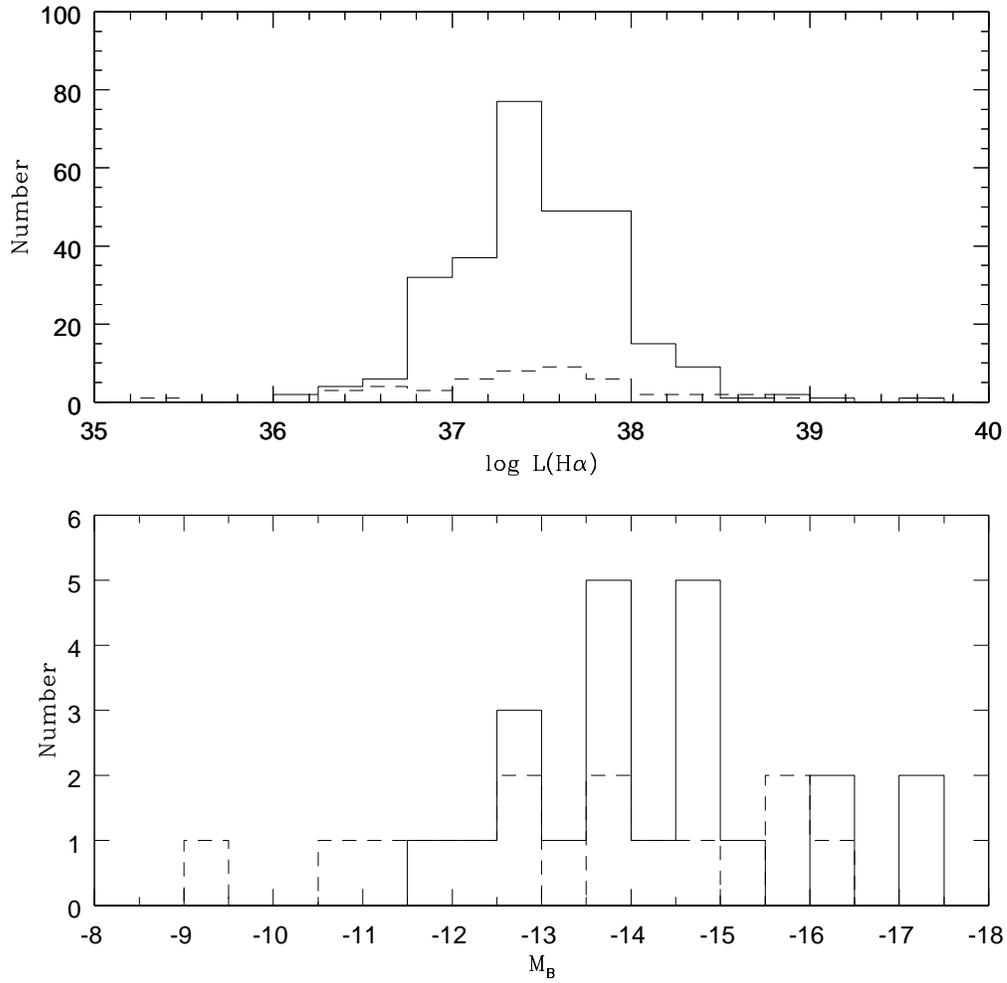}
\figcaption{Top: histogram of the HII region luminosities for the Centaurus~A Group
dIs. The dashed line histogram is for the Sculptor Group dIs \citep{scm03a}.
Bottom: histogram of absolute B magnitudes for the Centaurus A Group and Sculptor Group dIs.
The full line is for the Centaurus~A Group, and dash for the Sculptor Group.
\label{fig2}}
\end {figure}

\clearpage 

\begin {figure}
\plotone{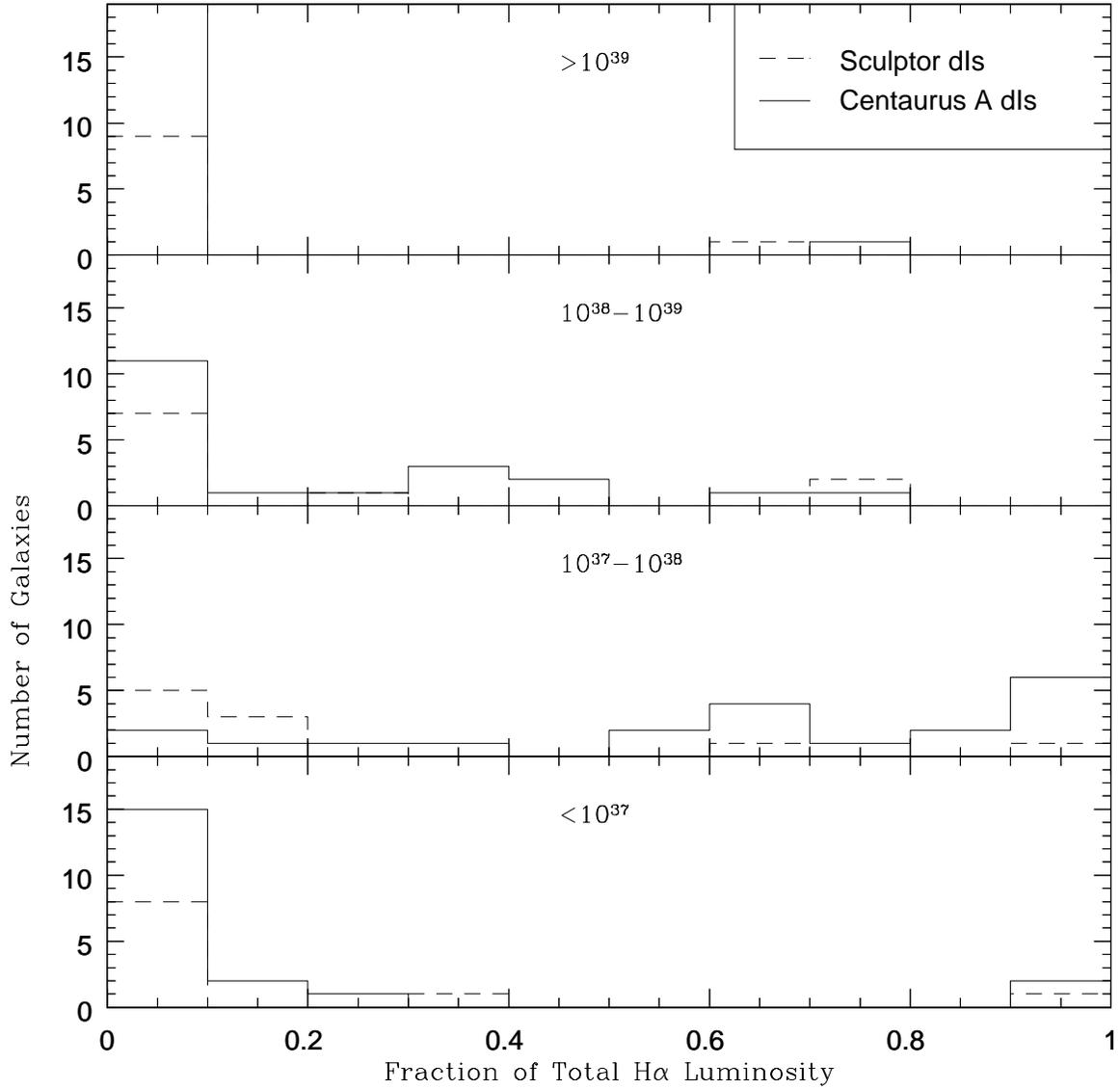}
\figcaption{
Fraction of the total H$\alpha$ luminosity in each galaxy that is contributed
by HII regions of a particular luminosity range, ranging from regions with
luminosities like that of the Orion nebula $\sim 10^{37}$ erg s$^{-1}$ 
(bottom) to supergiant HII regions with $L\geq 10^{39}$ erg s$^{-1}$ (top).
\label{fig3}}
\end{figure}

\clearpage 

\begin {figure}
\plotone{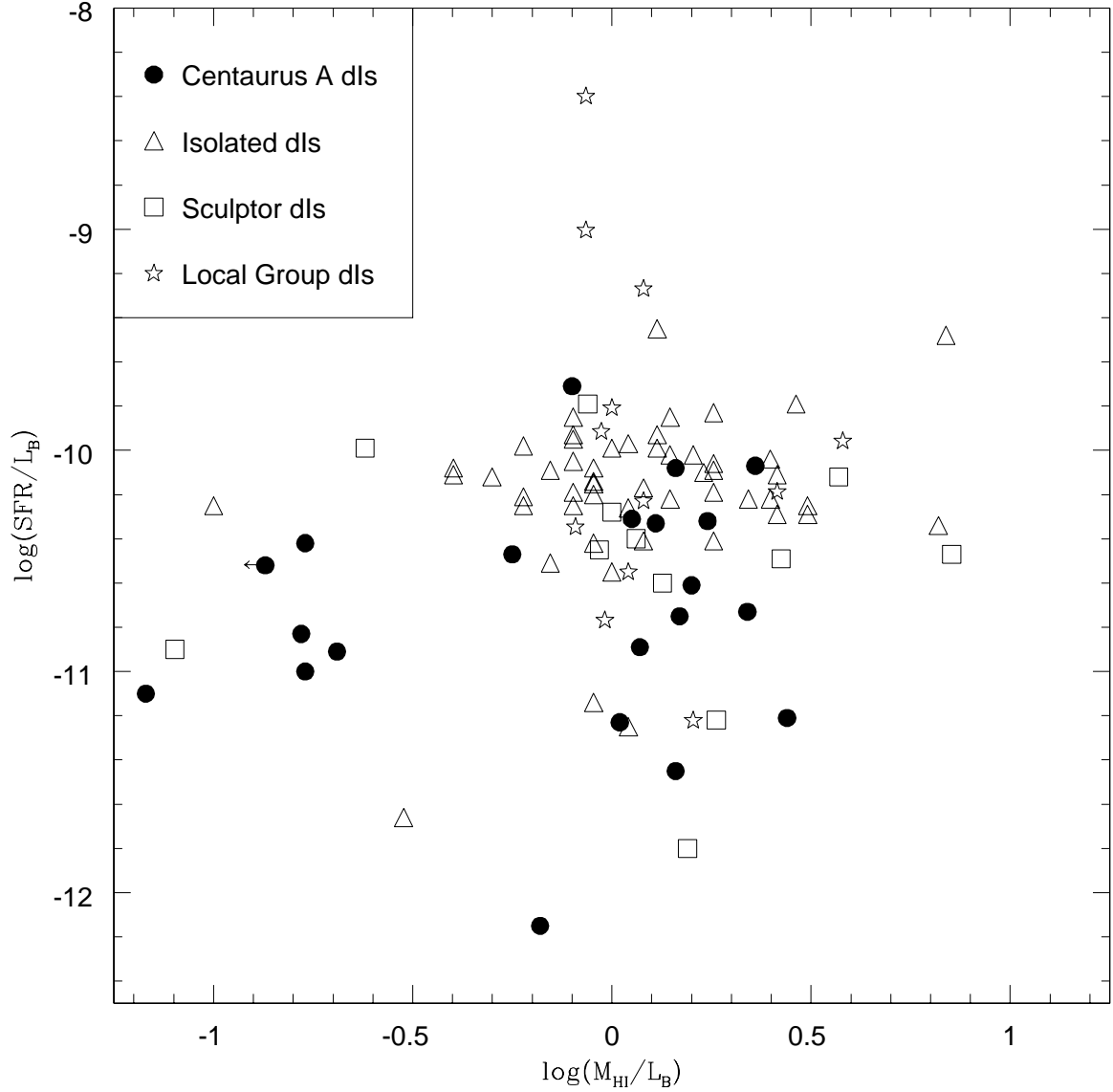}
\figcaption{
A comparison of SFR and gas mass normalized to the galaxy luminosity
for the Centaurus~A Group dIs and
three comparison groups:
the Local Group dIs \citep[from][]{m98}, the Sculptor Group dIs 
(from Skillman \etal 2003), and the
isolated dIs of van Zee (2001).
\label{fig4}}
\end{figure}

\clearpage 

\begin {figure}
\plotone{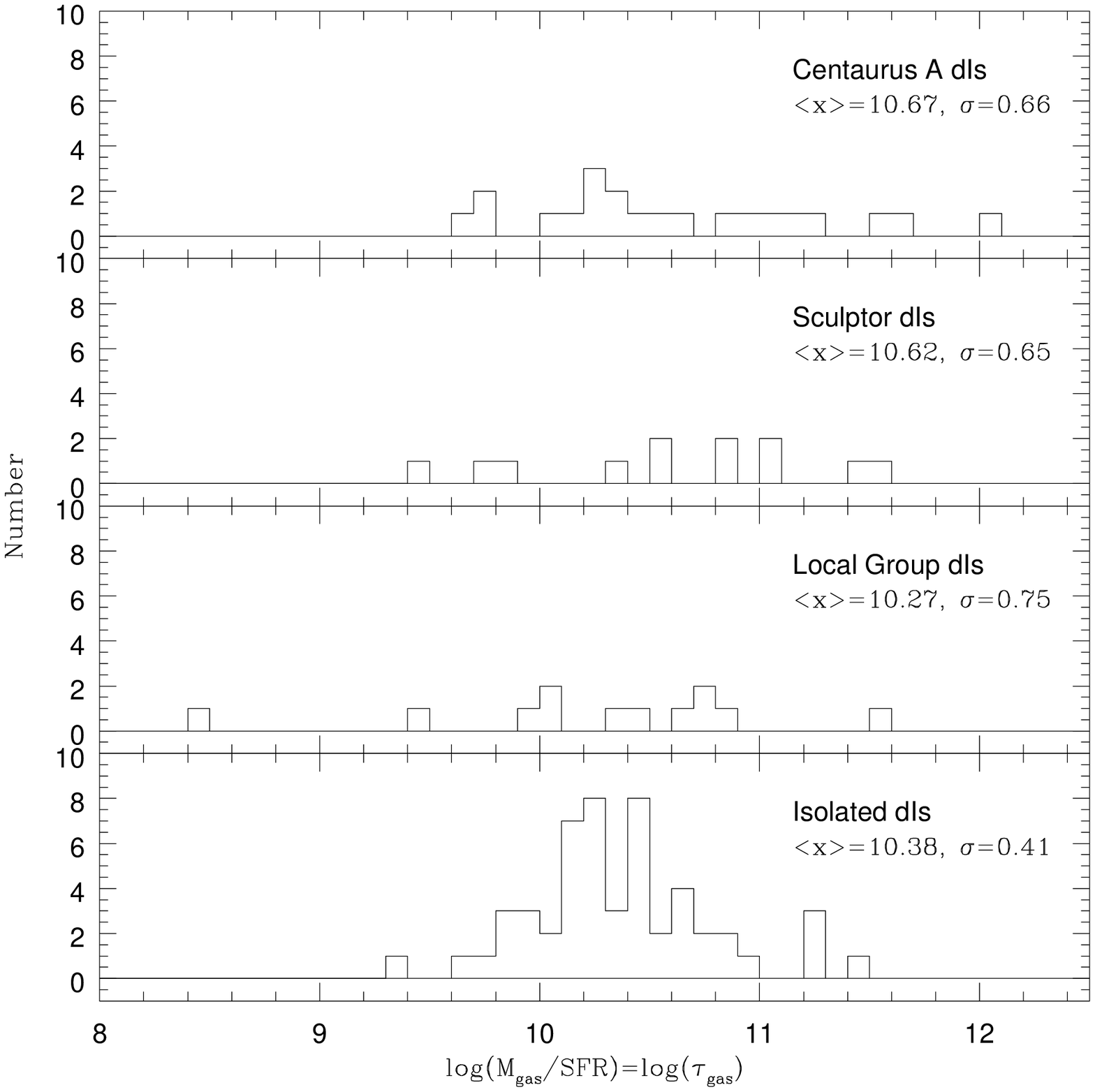}
\figcaption{
A comparison of the ratio of the gas mass to the current star formation rate
($=$ $\tau_{gas}$) for the Centaurus~A Group dIs and
three comparison groups: the Sculptor Group dIs (Skillman \etal 2003), 
the Local Group dIs \citep[from][]{m98}, and the 
isolated dIs of van Zee (2001).
For each sample, the mean value and the standard deviation in the sample is given.
\label{fig5}}
\end{figure}

\clearpage

\begin {figure}
\plotone{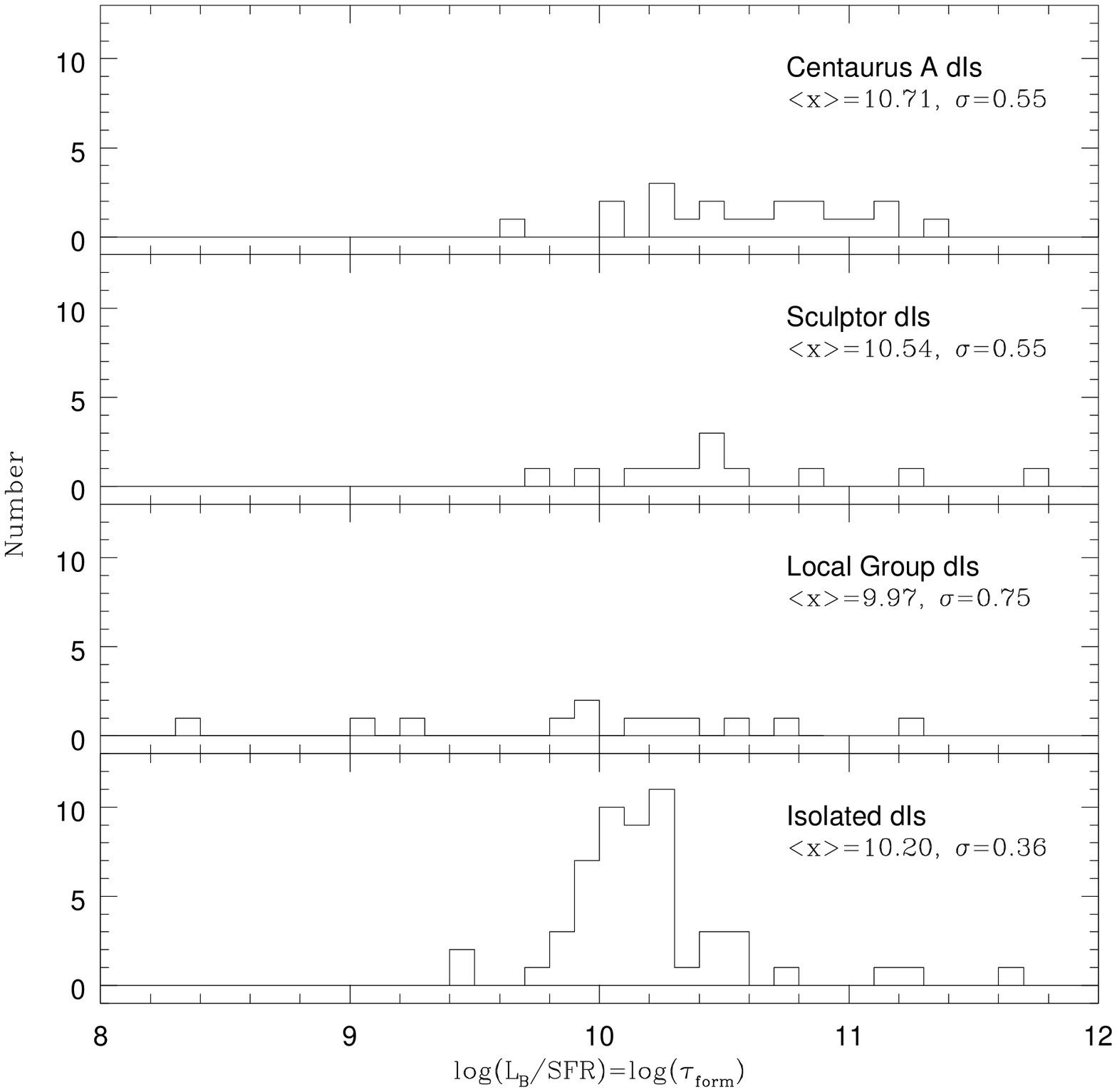}
\figcaption{
A comparison of the ratio of the luminosity to the current star formation rate
($=$ $\tau_{form}$) for the Centaurus~A Group dIs and
three comparison groups: the Sculptor Group dIs (Skillman \etal 2003),
the Local Group dIs \citep[from][]{m98}, and the 
isolated dIs of van Zee (2001).
For each sample, the mean value and the standard deviation in the sample is given.
\label{fig6}}
\end{figure}

\clearpage 

\begin {figure}
\plotone{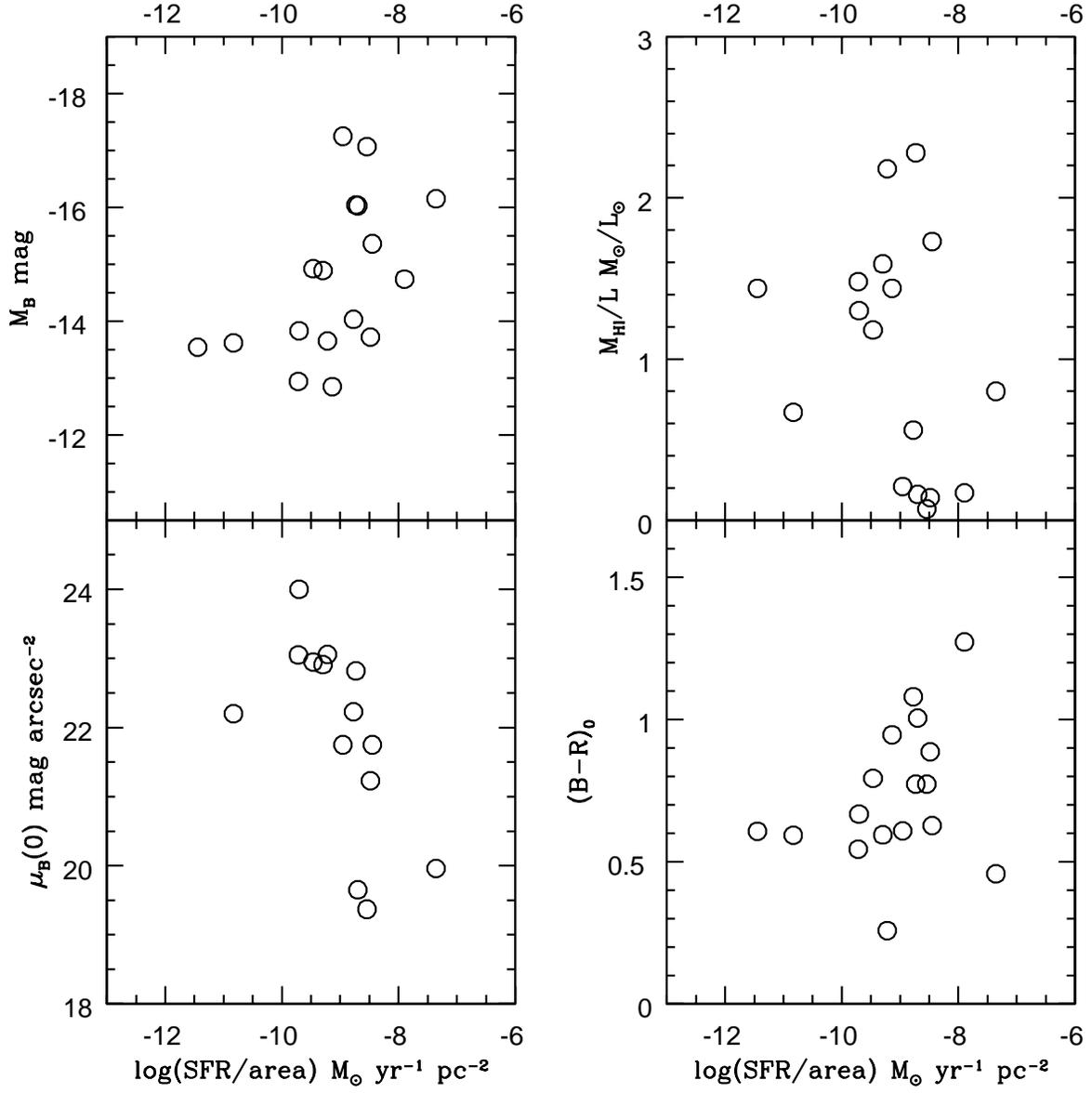}
\figcaption{
The star formation rates per unit area plotted against various global parameters: 
the absolute B magnitude $M_B$; the B central surface brightness $\mu _0$; the
HI mass to luminosity ratio $M_{HI}/L_B$; the color B-R.
\label{fig7}}
\end{figure}

\clearpage 

\begin {figure}
\plotone{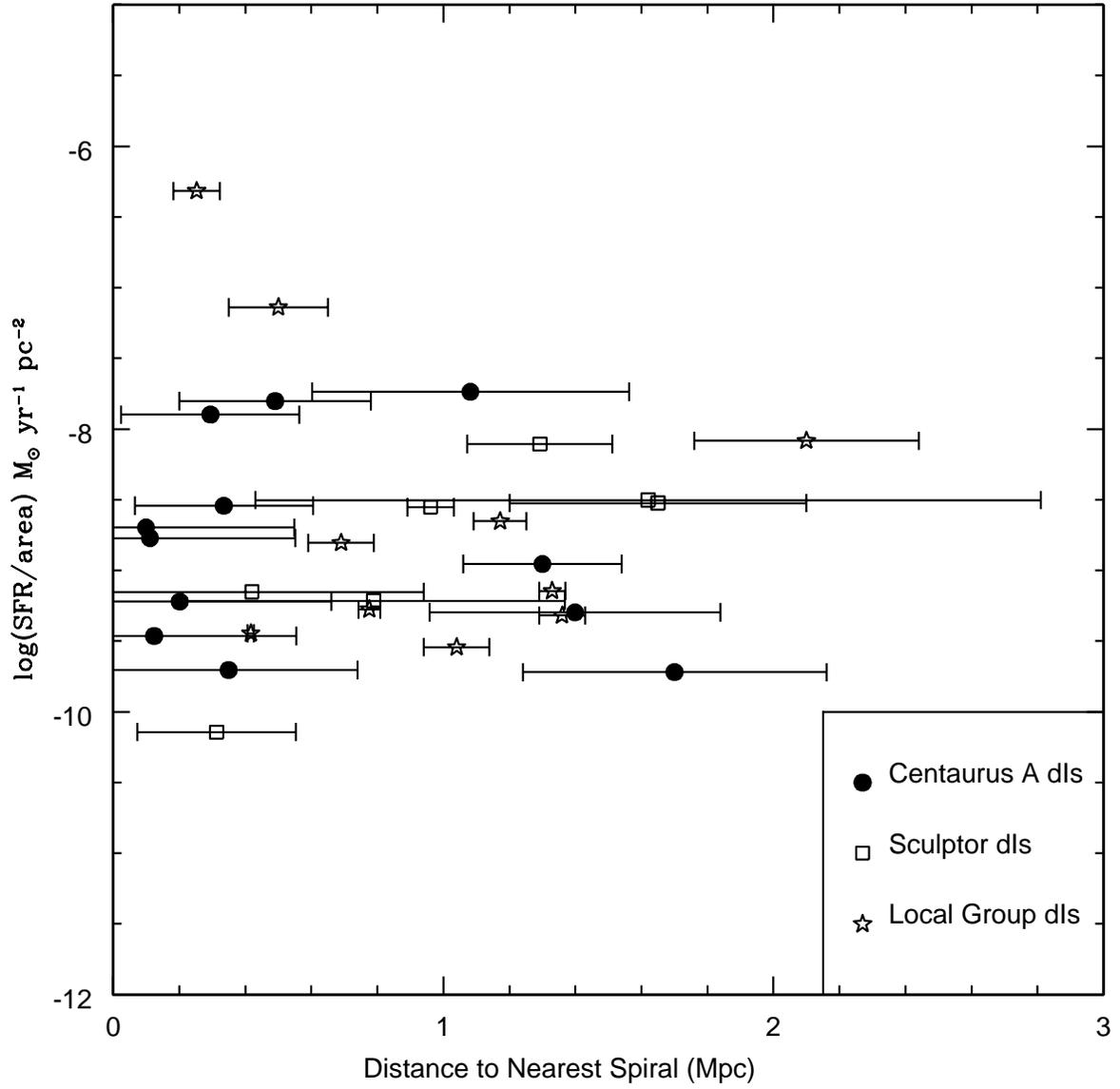}
\figcaption{
The star formation rates per unit area versus the distance to the nearest
large galaxy in the group. As above, dots are for Centaurus~A Group dIs, squares
for Sculptor Group dIs and stars for Local Group dIs.
\label{fig8}}
\end{figure}

\clearpage 

\begin {figure}
\plotone{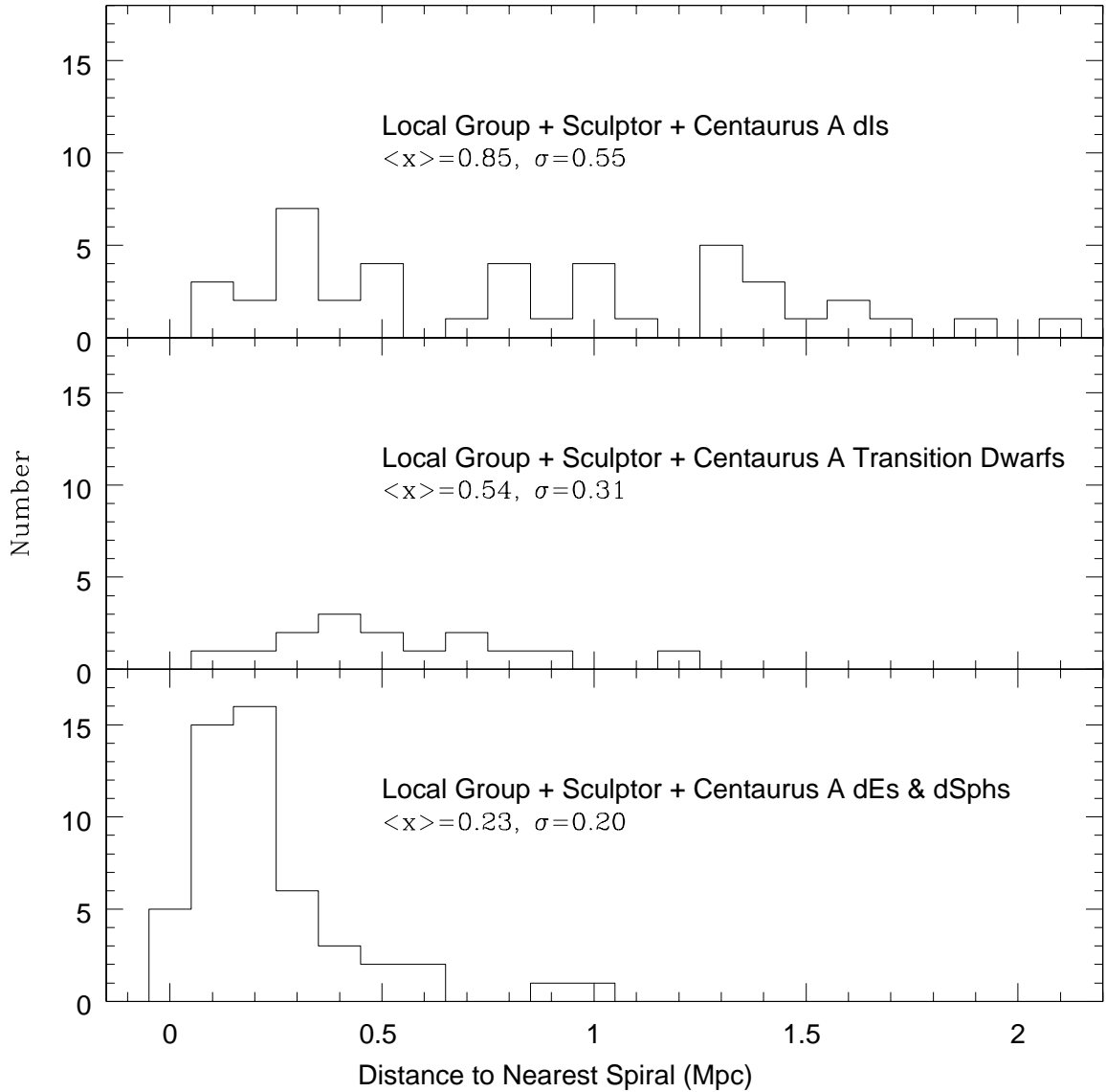}
\figcaption{
Histogram comparison of the distances to the nearest large galaxy of the group
for the dI, transition, and dE galaxies in the Local Group, Sculptor
Group and Centaurus~A Group. For each sample the mean value and the standard
deviation in the sample is given. On average the dIs lie at preferentially
larger distances from the main galaxies of the group than the transition 
dwarfs, which are themselves at larger distances than the dEs.
\label{fig9}}
\end{figure}

\clearpage

\begin {figure}
\plotone{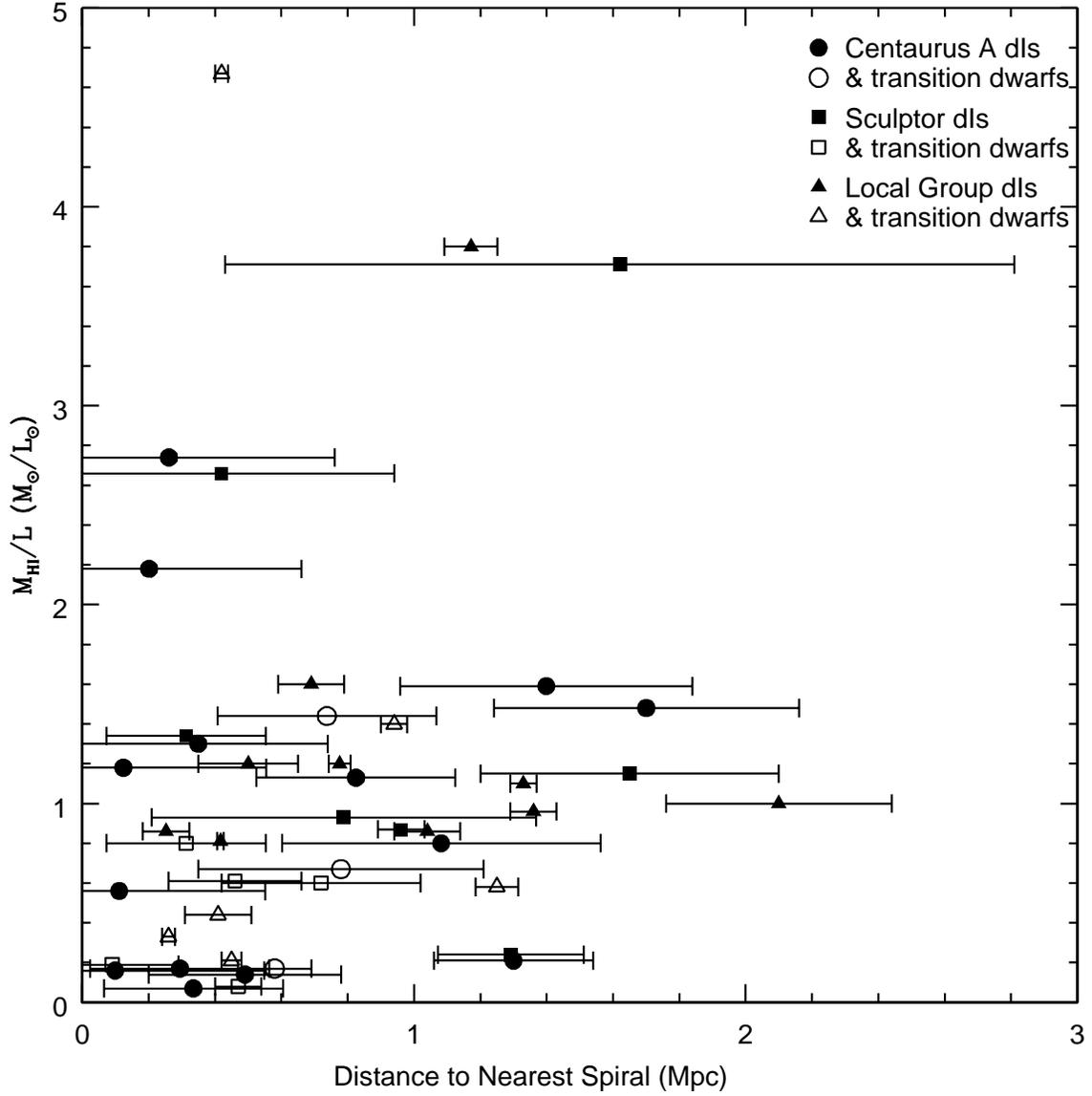}
\figcaption{
HI mass to light ratio M$_{HI}/L$ is plotted
versus the distance to the nearest
large galaxy in the group. Full dots are for Centaurus~A Group dIs, full squares
for Sculptor Group dIs and full triangles for Local group dIs, while open symbols
are for the transition dwarfs for each group respectively. The dIs closest to
their primary galaxy are not less HI-rich than the dIs further out.
\label{fig10}}
\end{figure}

\clearpage


\newpage

\begin{thebibliography}{}


\bibitem[Aguerri et al.(2005)]{aivms05} 
Aguerri, J.~A.~L., Iglesias-P{\'a}ramo, J., V{\'{\i}}lchez, J.~M., 
Mu{\~n}oz-Tu{\~n}{\'o}n, C., \& S{\'a}nchez-Janssen, R.\ 2005, \aj, 130, 475 

\bibitem[Armandroff et al.(1999)]{ajd99}
Armandroff, T.~E., Jacoby, G.~H., \& Davies, J.~E.\ 1999, \aj, 118, 1220 

\bibitem[Babul \& Rees(1992)]{br92}
Babul, A.~\& Rees, M.~J.\ 1992, \mnras, 255, 346 

\bibitem[Balogh et al.(1997)]{bal97} 
Balogh, M., Morris, S., Yee, H., Carlberg, R., Ellingson, E.\ 1997, \apj, 488, L75 

\bibitem[Banks et al.(1999)]{b99} 
Banks, G.~D., et al.\ 1999, \apj, 524, 612 

\bibitem[Barkana \& Loeb(1999)]{bl99}
Barkana, R., \& Loeb, A. 1999, ApJ, 523, 54

\bibitem[Barnes et al.(2001)]{hipass01}
Barnes, D.~G.~et al.\ 2001, \mnras, 322, 486 

\bibitem[Beaulieu et al.(2006)]{b06} 
Beaulieu, S., Freeman, K.C., Carignan, C., \&  Lockman, F.J.\ 2006, AJ, 131, 325 

\bibitem[Begum \& Chengalur(2005)]{bc05} 
Begum, A., \& Chengalur, J.~N.\ 2005, \mnras, 362, 609 

\bibitem[Bell (2003)]{b03}
Bell, E.F.\ 2003, ApJ, 586, 794

\bibitem[Binggeli et al.(1990)]{bts90}
Binggeli, B., Tarenghi, M., \& Sandage, A. 1990, A\&A, 228, 42

\bibitem[Blitz \& Robishaw(2000)]{br00}
Blitz, L.~\& Robishaw, T.\ 2000, \apj, 541, 675 

\bibitem[Bomans et al.(1997)]{bch97}
Bomans, D.J., Chu, Y.H., \& Hopp, U. 1997, \aj , 113, 1678

\bibitem[Bomans \& Grant(1998)]{bg98}
Bomans, D.J., \& Grant, M.-B. 1998, Astron.\ Nach., 319, 26

\bibitem[Boissier et al.(2007)]{boi07} 
Boissier, S., Gil de Paz, A., Boselli, A. et al. \ 2007, \apj, 173, 524 

\bibitem[Bouchard et al.(2003)]{bcm03} 
Bouchard, A., Carignan, C., \& Mashchenko, S.\ 2003, \aj, 126, 1295 

\bibitem[Bouchard et al.(2004)]{bdj04} 
Bouchard, A., Da Costa, G.~S., \& Jerjen, H.\ 2004, \pasp, 116, 1031 

\bibitem[Bouchard et al.(2005)]{bjdo05} 
Bouchard, A., Jerjen, H., Da Costa, G.~S., \& Ott, J.\ 2005, \aj, 130, 2058 

\bibitem[Bouchard et al.(2007)]{bjdo07} 
Bouchard, A., Jerjen, H., Da Costa, G.~S., \& Ott, J.\ 2007, \aj, 133, 261 

\bibitem[Bouchard et al.(2009)]{bou09} 
Bouchard, A., Da Costa, G.~S., Jerjen, H.\ 2009, AJ, 137, 3038

\bibitem[Brosch et al.(1998)]{bha98}
Brosch, N., Heller, A., \& Almoznino, E. 1998, \mnras , 300, 1091

\bibitem[Bullock et al.(2000)]{bkw00}
Bullock, J.S., Kravtsov, A.V., \& Weinberg, D.H. 2000, ApJ, 539, 517

\bibitem[Cannon et al.(2003)]{cdsbcm03} 
Cannon, J.~M., Dohm-Palmer, R.~C., Skillman, E.~D., Bomans, D.~J., 
C{\^o}t{\'e}, S., \& Miller, B.~W.\ 2003, \aj, 126, 2806 

\bibitem[Carraro et al.(2001)]{ccgl01}
Carraro, G., Chiosi, C., Girardi, L., \& Lia, C. 2001, \mnras, 327, 69 

\bibitem[Calzetti et al.(1999)]{cal99}
Calzetti, D., Conselice, C.J., Gallagher, J.S. \& Kinney, A.L. 1999, AJ, 118, 797

\bibitem[Carignan et al.(1991)]{cdc91}
Carignan, C., Demers, S. \& C\^ot\'e, S. 1991, ApJ, 381, L13

\bibitem[Chung et al.(2007)]{cvg07}
Chung, A., van GOrkom, J.H., Kenney, J. \& Vollmer, B. 2007, ApJ, 659, L115 

\bibitem[Conselice et al.(2003)]{cogw03} 
Conselice, C.~J., O'Neil, K., Gallagher, J.~S., \& Wyse, R.~F.~G.\ 2003, \apj, 591, 167 

\bibitem[Conselice(2006)]{c06} 
Conselice, C.~J.\ 2006, ArXiv Astrophysics e-prints, arXiv:astro-ph/0605531 

\bibitem[C\^ot\'e(1995)]{c95}
C\^ot\'e, S. 1995, Ph.D.\ Thesis, Australian National University

\bibitem[C\^ot\'e et al.(2000)]{ccf00}
C\^ot\'e, S., Carignan, C., \& Freeman, K.C. 2000, \aj , 120, 3027

\bibitem[C\^ot\'e et al.(1997)] {cfcq97}
C\^ot\'e, S., Freeman, K. C., Carignan, C., \& Quinn, P. 1997, AJ, 114, 1313

bibitem[C\^ot\'e et al.(2006)] {cpf06}
C\^ot\'e, P., Piatek, S., Ferrarese, L.~et al. 2006, ApJS, 165, 57

\bibitem[de Blok et al.(2002)]{dzdbf02} de Blok, W.~J.~G., 
Zwaan, M.~A., Dijkstra, M., Briggs, F.~H., \& Freeman, K.~C.\ 2002, \aap, 
382, 43 

\bibitem[da Costa et al.(2007)] {djb08}
Da Costa, G.~S., Jerjen, H., \& Bouchard, A.\ 2007, ArXiv Astrophysics e-prints, arXiv:astro-ph/0710.1420

\bibitem[Dav\'e et al.(2001)] {dave2001}
Dav\'e, R. et al.\ 2001, ApJ, 552, 473

\bibitem[Davidge (2008)] {da08}
Davidge, T.J.\ 2008, AJ, 135, 1636

\bibitem[De Rijcke et al.(2004)]{ddzh04} 
De Rijcke, S., Dejonghe, H., Zeilinger, W.~W., \& Hau, G.~K.~T.\ 2004, \aap, 426, 53 

\bibitem[de Vaucouleurs(1958)]{d58}
de Vaucouleurs, G. 1958, AJ, 63, 253

\bibitem[de Vaucouleurs(1975)]{d75}
de Vaucouleurs, G. 1975, in Stars and Stellar Systems 9, Galaxies
and the Universe, ed.\ A.\ Sandage, M.\ Sandage, \& J.\ Kristian
(Chicago: Univ.\ Chicago Press), 557

\bibitem[de Vaucouleurs(1979)]{deV79}
de Vaucouleurs, G. 1979, AJ, 84, 1270

\bibitem[de Vaucouleurs(1991)]{rc3}
de Vaucouleurs, G., de Vaucouleurs, A., Corwin, H. G. Jr.,
Buta, R. J., Paturel, G., \& Foqu\'e, P. 1991, Third Reference
Catalog of Bright Galaxies, (New York: Springer) (RC3)

\bibitem[Dohm-Palmer et al.(1997)]{dp97a}
Dohm-Palmer, R.~C.~et al.\ 1997, \aj, 114, 2527 

\bibitem[Dohm-Palmer et al.(1998)]{dp98}
Dohm-Palmer, R.~C.~et al.\ 1998, \aj, 116, 1227 

\bibitem[Done et al.(1996)]{do96}
Done, C., Madejski, G.M., \& Smith, D.A. \ 1996, \apj, 463, L63 

\bibitem[Efstathiou(1992)]{e92}
Efstathiou, G. 1992, \mnras, 256, 43P 

\bibitem[Elmegreen(1997)]{e97} 
Elmegreen, B.~G.\ 1997, \apj, 477, 196 

\bibitem[Ferguson(2002)]{f02} 
Ferguson, A.~M.~N.\ 2002, \apss, 281, 119 

\bibitem[Ferguson et al.(1996)]{fwgh96}
Ferguson, A.~M.~N., Wyse, R.~F.~G., Gallagher, J.~S., \& Hunter, D.~A.\ 
1996, \aj, 111, 2265 

\bibitem[Gallagher \& Hunter(1987)]{gh87} 
Gallagher, J.~S., III, \& Hunter, D.~A.\ 1987, \aj, 94, 43 

\bibitem[Gallagher et al.(2003)]{gmrgs03} 
Gallagher, J.~S., Madsen, G.~J., Reynolds, R.~J., Grebel, E.~K., \& 
    Smecker-Hane, T.~A.\ 2003, \apj, 588, 326 

\bibitem[Gallagher et al.(1998)]{gtdschsm98}
Gallagher, J. S., Tolstoy, E., Dohm-Palmer, R. C., Skillman, E. D., Cole, A., 
Hoessel, J., Saha, A., \& Mateo, M. 1998, AJ, 115, 1869

\bibitem[Gallagher (2005)]{ga05}
Gallagher, J. S. in Starbursts: From 30 Doradus to Lyman Break Galaxies,
eds. R. de Grijs \& R. Gonzalez Delgado (Dordrecht: Sringer), 11


\bibitem[Gallart et al.(2001)]{gmgm01}
Gallart, C., Martinez-Delgado, D., Gomez-Flechoso, M.A., Mateo, M. 2001, AJ, 121, 2572

\bibitem[Gavazzi et al.(1998)]{g98} 
Gavazzi, G., Catinella, B., Carrasco, L., Boselli, A., \& Contursi, A.\ 
1998, \aj, 115, 1745 

\bibitem[Gavazzi et al.(2002)]{g02} 
Gavazzi, G., Boselli, A., Pedotti, P., Gallazzi, A., \& Carrasco, L.\ 
2002, \aap, 396, 449 

\bibitem[Geha et al.(2006)]{ggrc06} 
Geha, M., Guhathakurta, P., Rich, R.~M., \& Cooper, M.~C.\ 2006, \aj, 131, 332 

\bibitem[Gnedin (2000)]{g00}
Gnedin, N. 2000, ApJ, 535, L75

\bibitem[Giuricin et al.(2000)]{gmcp00}
Giuricin, G., Marinoni, C., Ceriani, L., \& Pisani, A. 2000, ApJ, 543, 178

\bibitem[Grebel et al.(2003)]{ggh03} 
Grebel, E.~K., Gallagher, J.~S., III, \& Harbeck, D.\ 2003, \aj, 125, 1926

\bibitem[Grossi et al.(2007)]{g07} 
Grossi, M., Disney, M.~J., Pritzl, B.~J., Knezek, P.~M., Gallagher, J.~S., 
    Minchin, R.~F., \& Freeman, K.~C.\ 2007, \mnras, 374, 107 

\bibitem[Gunn \& Gott(1972)]{gg72} 
Gunn, J.~E., \& Gott, J.~R.~I.\ 1972, \apj, 176, 1 

\bibitem[Heisler et al.(1997)]{hhmh97}
Heisler, C.A., Hill, T.L., McCall, M.L., Hunstead, R.W. 1997, \mnras , 285, 374

\bibitem[Hirashita (2000)]{hi00}
Hirashita, H. 2000, \pasj , 52, 107

\bibitem[Hodge (1993)]{h93}
Hodge, P. 1993, in Star Formation, Galaxies, and the Interstellar Medium,
   eds.\ J.\ Franco, F.\ Ferrini, \& G.\ Tenorio-Tagle, Cambridge University 
   Press, 294

\bibitem[Holtzman et al.(2000)]{hsg00}
Holtzman, J.~A., Smith, G.~H., \& Grillmair, C.\ 2000, \aj, 120, 3060 

\bibitem[Hoversten \& Glazebrook(2008)]{hg08}
Hoversten, E.A. \& Glazebrook, K. \ 2008, ApJ, 675, 163 

\bibitem[Huchtmeier et al.(2000)]{hke00} 
Huchtmeier, W.~K., Karachentsev, I.~D., Karachentseva, V.~E., \& 
   Ehle, M.\ 2000, \aaps, 141, 469 

\bibitem[Huchtmeier et al.(2005)]{hkp05} 
Huchtmeier, W.~K., Krishna, G., \& Petrosian, A.\ 2005, \aap, 434, 887 

\bibitem[Hunter et al.(1982)]{hgr82} 
Hunter, D.~A., Gallagher, J.~S., \& Rautenkrantz, D.\ 1982, ApJS, 49, 53


\bibitem[Hunter \& Gallagher(1986)]{hg86} 
Hunter, D.~A., \& Gallagher, J.~S., III 1986, \pasp, 98, 5 

\bibitem[Hunter et al.(1993)]{hhg93}  
Hunter, D. A., Hawley, W. N., \& Gallagher, J. S. 1993, AJ, 106, 1797

\bibitem[Hunter \& Elmegreen(2004)]{he04}  
Hunter, D. A., \& Elmegreen, B.G. 2004, AJ, 128, 2170

\bibitem[Iglesias-P{\'a}ramo \& V{\'{\i}}lchez(1999)]{iv99} 
Iglesias-P{\'a}ramo, J., \& V{\'{\i}}lchez, J.~M.\ 1999, \apj, 518, 94 

\bibitem[Irwin \& Tolstoy (2002)]{it02} 
Irwin, M.~\& Tolstoy, E.\ 2002, \mnras, 336, 643 

\bibitem[Irwin et al.(2007)]{ibe07} 
Irwin, M. et al.\ 2007, ApJ, 656, L13 

\bibitem[Jerjen et al.(1998)]{jfb98}
Jerjen, H., Freeman, K. C., \& Binggeli, B. 1998, AJ, 116, 2873

\bibitem[Jerjen (2000)]{j00}
Jerjen, H. 2000, AJ, 119, 166 
 
\bibitem[Jerjen \& Rejkuba (2000)]{jr01}
Jerjen, H., \& Rejkuba, M. 2000, A\&A, 371, 487 

\bibitem[Kaisin et al.(2007)]{kai07} 
Kaisin, S., Kasparova, A., Knyazev, A., Karachentsev, I.\ 2007, AstL, 33, 283 

\bibitem[Karachentsev(2005)]{k05} 
Karachentsev, I.~D.\ 2005, \aj, 129, 178 

\bibitem[Karachentsev \& Kaisin(2007)]{kk07} 
Karachentsev, I.~D., \& Kaisin, S.~S.\ 2007, \aj, 133, 1883 

\bibitem[Karachentsev et al.(2000)]{k00}
Karachentsev, I.~D.~et al.\ 2000, \apj, 542, 128 

\bibitem[Karachentsev et al.(2002)]{k02} 
Karachentsev, I.~D., et al.\ 2002, \aap, 385, 21 

\bibitem[Karachentsev et al.(2007)]{k07} 
Karachentsev, I.~D., et al.\ 2007, \aj, 133, 504 

\bibitem[Karachentseva \& Karachentsev (1998)]{kk98} 
Karachentseva, V.E, \& Karachentsev, I.D.\ 1998, A\&AS, 127, 409 

\bibitem[Karachentseva \& Karachentsev(2000)]{kk00} 
Karachentseva, V.~E., \& Karachentsev, I.~D.\ 2000, \aaps, 146, 359 

\bibitem[Kennicutt(1983)]{k83}
Kennicutt, R.C. Jr. 1983, \apj , 272, 54

\bibitem[Kennicutt(1984)]{k84}
Kennicutt, R.C. Jr. 1984, \apj , 287, 116

\bibitem[Kennicutt(1989)]{k89} 
Kennicutt, R.~C., Jr.\ 1989, \apj, 344, 685 

\bibitem[Kennicutt(1998)]{k98}
Kennicutt, R.C. Jr. 1998, \apj , 498, 541

\bibitem[Kennicutt \& Hodge(1986)]{kh86}
Kennicutt, R.C. Jr., \& Hodge, P.W. 1986, \apj , 306, 130

\bibitem[Kennicutt \& Skillman(2001)]{ks01}
Kennicutt, R.C. Jr., \& Skillman, E.D. 2001, \aj , 121, 1461 

\bibitem[Kennicutt et al.(1994)]{ktc94}
Kennicutt, R.C. Jr., Tamblyn, P., \& Congdon, C.W. 1994, \apj , 435, 22

\bibitem[Kennicutt et al.(2008)]{kl08}
Kennicutt, R.C. Jr., Lee, J., Funes, J., Sakai, S. \& Akiyama, S. 2008, ApJS, 178, 247

\bibitem[Kewley et al.(2001)]{khdl01} 
Kewley, L.~J., Heisler, C.~A., Dopita, M.~A., \& Lumsden, S.\ 2001, \apjs, 132, 37 

\bibitem[Klypin et al.(1999)]{kkvp99}
Klypin, A., Kravtsov, A. V., Valenzuela, O,, \& Prada, F. 1999, ApJ, 522, 82

\bibitem[Knezek et al.(1999)]{ksg99}
Knezek, P. M., Sembach, K. R., \& Gallagher, J. S., III 1999, ApJ, 514, 119

\bibitem[Koopmann \& Kenney (2006)]{kk06}
Koopmann, R. \& Kenney, J. 2006, ApJS, 162, 97

\bibitem[Larson et al.(1980)]{ltc80} 
Larson, R.~B., Tinsley, B.~M., \& Caldwell, C.~N.\ 1980, \apj, 237, 692 

\bibitem[Lauberts(1984)]{l84} 
Lauberts, A.\ 1984, \aaps, 58, 249 

\bibitem[Lauberts \& Valentijn(1989)]{lv89} 
Lauberts, A., \& Valentijn, E.~A.\ 1989,
The Surface Photometry Catalogue of the ESO‐Uppsala Galaxies,
Garching: European Southern Observatory  

\bibitem[Lee et al.(2006)]{lea06} 
Lee, H., Skillman, E.~D., Cannon, J.~M., Jackson, D.~C., Gehrz, R.~D., 
Polomski, E.~F., \& Woodward, C.~E.\ 2006, \apj, 647, 970 

\bibitem[Lee (2006)]{lee06} 
Lee, J.C.\ 2006, PhD Thesis, University of Arizona

\bibitem[Lee (2009)]{lee09} 
Lee, J.C., Kennicutt, R.C., Funes, J.G., Sakai, S., Akiyama, S.\ 2009, ApJ, 692, 1305

\bibitem[Lisker et al.(2006)]{lgwg06} 
Lisker, T., Glatt, K., Westera, P., \& Grebel, E.~K.\ 2006, \aj, 132, 2432

\bibitem[Lo et al.(1993)]{lsy93} 
Lo, K.~Y., Sargent, W.~L.~W., \& Young, K.\ 1993, \aj, 106, 507 

\bibitem[Marlowe et al.(1997)]{mmhs97}
Marlowe, A. T., Meurer, G. R., Heckman, T. M., \& Schommer, R. 1997, 
\apjs , 112, 285

\bibitem[Mashchenko et al.(2004)]{mcb04} 
Mashchenko, S., Carignan, C., \& Bouchard, A.\ 2004, \mnras, 352, 168 

\bibitem[Mateo(1998)]{m98}
Mateo, M. 1998, ARA\&A, 36, 435

\bibitem[Mayer et al.(2001a)]{mgcmqwsl01a}
Mayer, L., Governato, F., Colpi, M., Moore, B., Quinn, T., Wadsley, J., 
Stadel, J., \& Lake, G. 2001a, ApJ, 547, L123

\bibitem[Mayer et al.(2001b)]{mgcmqwsl01b}
Mayer, L., Governato, F., Colpi, M., Moore, B., Quinn, T., Wadsley, J., 
Stadel, J., \& Lake, G. 2001b, \apj, 559, 754 

\bibitem[Mayer et al.(2006)]{mmwsm06} 
Mayer, L., Mastropietro, C., Wadsley, J., Stadel, J., \& 
   Moore, B.\ 2006, \mnras, 369, 1021

\bibitem[Mayer et al.(2007)]{mkmw07} 
Mayer, L., Kazantzidis, S., Mastropietro, C., \& Wadsley, J.\ 2007, \nat, 445, 738 

\bibitem[Meurer et al.(2006)]{meu06} 
Meurer, G., Hanish, D., Ferguson, H. et al \ 2006, ApJS, 165, 307 

\bibitem[Miller(1994)]{m94}
Miller, B. W. 1994, Ph.D.\ Thesis, University of Washington

\bibitem[Miller(1996)]{m96}
Miller, B. W. 1996, AJ, 112, 991 

\bibitem[Miller et al.(2001)]{mdlkh01}
Miller, B.~W., Dolphin, A.~E., Lee, M.~G., Kim, S.~C., \& Hodge, P. 2001
\apj , 562, 713

\bibitem[Miller \& Hodge(1994)]{mh94}
Miller, B.W., \& Hodge, P. 1994, \apj , 427, 656

\bibitem[Minchin et al.(2003)]{m03} 
Minchin, R.~F., et al.\ 2003, \mnras, 346, 787 

\bibitem[Moore et al.(1996)]{mkldo96} 
Moore, B., Katz, N., Lake, G., Dressler, A., \& Oemler, A.\ 1996, \nat, 379, 613 

\bibitem[Moore et al.(1999)]{mgglqst99}
Moore, B., Ghigna, S., Governato, F., Lake, G., Quinn, T., Stadel, J., \&
Tozzi, P. 1999, ApJ, 524, L19

\bibitem[Nicastro et al.(2002)]{ni02}
Nicastro, F. et al. 2002, ApJ, 573, 157

\bibitem[Normandeau(1996)]{ntd96}
Normandeau, M., Taylor, A.R., \& Dewdney, P.E. 1996, Nature, 380, 687

\bibitem[Oosterloo et al.(1996)]{ods96}
Oosterloo, T., Da Costa, G.S., \& Staveley-Smith, L. 1996, AJ, 112, 1969

\bibitem[Peebles(1989)]{p89} 
Peebles, P.~J.~E.\ 1989, \apjl, 344, L53 

\bibitem[Perez-Gonzalez et al.(2003)]{pg03} 
Perez-Gonzalez, P., Zamorano, J., Gallego, J. et al.\ 2003, \apj, 591, 827 

\bibitem[Phillips et al.(1986)]{ph86} 
Phillips, M.M., Jenkins, C.R., Dopita, M.A., Sadler, E.M. \& Binette, L.\ 1986, AJ, 91, 1062

\bibitem[Press et al.(1992)]{ptvf92}
Press, W. H., Teukolsky, S. A., Vetterling, W. T., \& Flannery, B. P. 1992,
Numerical Recipes in Fortran, Cambridge University Press

\bibitem[Prugniel et al.(1993)]{pbka93} 
Prugniel, P., Bica, E., Klotz, A., \& Alloin, D.\ 1993, \aaps, 98, 229 

\bibitem[Prugniel \& Heraudeau(1998)]{ph98} 
Prugniel, P., \& Heraudeau, P.\ 1998, \aaps, 128, 299 

\bibitem[Puche \& Carignan(1988)]{pc88}
Puche, D., \& Carignan, C. 1988, AJ, 95, 1025

\bibitem[Quinn et al.(1996)]{qke96}
Quinn, T., Katz, N., \& Efstathiou, G.\ 1996, \mnras, 278, L49 

\bibitem[Richer et al.(2001)]{rbbmlkgkrr01}
Richer, M.~G.~et al.\ 2001, \aap, 370, 34 

\bibitem[Roberts(1963)]{r63}
Roberts, M.S. 1963, ARA\&A, 1, 149

\bibitem[Rumstay \& Kaufman(1983)]{rk83} 
Rumstay, K.~S., \& Kaufman, M.\ 1983, \apj, 274, 611 

\bibitem[Sadler(2001)]{s01} 
Sadler, E.~M.\ 2001, Gas and Galaxy Evolution, 
Eds. J. E. Hibbard, M. Rupen, and J. H. van Gorkom,
ASP Conference Proceedings, 240, 445 

\bibitem[Sandage \& Binggeli(1984)]{sb84} 
Sandage, A., \& Binggeli, B.\ 1984, \aj, 89, 919

\bibitem[Sandage \& Hoffman(1991)]{sh91}
Sandage, A., \& Hoffman, G.L. 1991, ApJ, 379, 45

\bibitem[Scalo(1986)]{s86}
Scalo, J.M. 1986, Fund.\ Cos.\ Phys., 11, 1

\bibitem[Schaerer et al.(1999)]{scp99} 
Schaerer, D., Contini, T., \& Pindao, M.\ 1999, \aaps, 136, 35 

\bibitem[Schlegel et al.(1998)]{sfd98}
Schlegel, D.J., Finkbeiner, D.P., \& Davis, M. 1998, \apj , 500, 525

\bibitem[Schaye(2004)]{s04} 
Schaye, J.\ 2004, \apj, 609, 667 

\bibitem[Sembach et al.(2003)]{sem03}
Sembach, K.R., et al. 2003, ApJS, 146, 165

\bibitem[Skillman(1996)]{s96}
Skillman, E.~D.\ 1996, ASP Conf.~Ser.~106: 
The Minnesota Lectures on Extragalactic Neutral Hydrogen, 208 

\bibitem[Skillman et al.(1997)]{sbk97}
Skillman, E. D., Bomans, D. J., \& Kobulnicky, H. A.  
1997, ApJ, 474, 205

\bibitem[Skillman et al.(2003a)]{scm03a} 
Skillman, E.~D., C{\^o}t{\'e}, S., \& Miller, B.~W.\ 2003, \aj, 125, 593

\bibitem[Skillman et al.(2003b)]{scm03b} 
Skillman, E.~D., C{\^o}t{\'e}, S., \& Miller, B.~W.\ 2003, \aj, 125, 610 

\bibitem[Skillman et al.(1988)]{s88} 
Skillman, E.~D., Terlevich, R., Teuben, P.~J., \& van Woerden, H.\ 1988, \aap, 198, 33 

\bibitem[St-Germain et al.(1999)]{scco99}
St-Germain, J., Carignan, C., C\^ot\'e, S., \& Oosterloo, T. 1999, AJ, 118, 1235

\bibitem[Strobel et al.(1991)]{shk91} 
Strobel, N.~V., Hodge, P., \& Kennicutt, R.~C., Jr.\ 1991, \apj, 383, 148 

\bibitem[Takei et al.(2007)]{t07} 
Takei, Y., Henry, P., Finoguenov, A. et al.\ 2007, ApJ, 655, 831 

\bibitem[Taylor et al.(1994)]{t94} 
Taylor, C.~L., Brinks, E., Pogge, R.~W., \& Skillman, E.~D.\ 1994, \aj, 107, 971 

\bibitem[Thomson(1992)]{t92} 
Thomson, R.~C.\ 1992, \mnras, 257, 689 

\bibitem[Toomre(1964)]{t64} 
Toomre, A.\ 1964, \apj, 139, 1217 

\bibitem[Tremonti et al.(2007)]{tre07} 
Tremonti, C.A., Lee, J.C., van Zee, L. et al. \ 2007, AAS, 211, 9503

\bibitem[Tully \& Fisher(1987)]{tf87}
Tully, R.B., \& Fisher, J.R. 1987, Nearby Galaxies Atlas, Cambridge
University Press
 
\bibitem[van den Bergh(1994a)]{vdb94a}
van den Bergh, S. 1994a, AJ, 107, 1328

\bibitem[van den Bergh(1994b)]{vdb94b}
van den Bergh, S. 1994b, ApJ, 428, 617

\bibitem[van den Bergh(2000)]{vdb00}
van den Bergh, S. 2000, PASP, 112, 529

\bibitem[van Zee(2000)]{vz00}
van Zee, L. 2000, \aj , 119, 2757

\bibitem[van Zee(2001)]{vz01}
van Zee, L. 2001, \aj , 121, 2003

\bibitem[van Zee et al.(1997a)]{vz97}
van Zee, L., Haynes, M. P., Salzer, J. J., Boriels, A. 1997, AJ, 113, 1618

\bibitem[van Zee et al.(1997b)]{vhs97}
van Zee, L., Haynes, M. P., \& Salzer, J. J. 1997, AJ, 114, 2479

\bibitem[Vorontsov-Vel'Yaminov \& Ivani{\v s}evi{\'c}(1974)]{vvi74} 
Vorontsov-Vel'Yaminov, B.~A., \& 
Ivani{\v s}evi{\'c}, G.\ 1974, Soviet Astronomy, 18, 174 

\bibitem[Whiting(1999)]{w99}
Whiting, A. B. 1999, AJ, 117, 202

\bibitem[Young \& Lo(1997)]{yl97}
Young, L.M., \& Lo, K.Y. 1997, ApJ, 490, 710
 
\bibitem[Youngblood \& Hunter(1999)]{yh99}
Youngblood, A.J., \& Hunter, D.A. 1999, \apj , 519, 55

\end{thebibliography}
\end{document}